\documentclass[aps,prd,twocolumn,reprint, longbibliography,  breaklinks=true, showkeys, floatfix, nofootinbib]{revtex4-2}
\usepackage[T1]{fontenc}
\usepackage[latin9]{inputenc}
\usepackage{amsmath}
\usepackage{amsthm}
\usepackage{amsfonts}
\usepackage{amssymb}
\usepackage{subfigure}
\usepackage{graphicx}

\usepackage{hyperref}
\pdfoutput=1
\hypersetup{colorlinks=true,citecolor=red,linkcolor=blue,urlcolor=blue,filecolor= green, breaklinks=true}

\begin{document}

\title{Compact regular objects from an electrified Tolman-like density: A new interior region for the Kerr-Newman spacetime}

\author{Marcos L. W. Basso} 

\author{Vilson T. Zanchin}
\affiliation{Centro de Ci\^encias Naturais e Humanas, Universidade Federal do ABC, Avenida dos Estados 5001, Santo Andr\'e, S\~ao Paulo, 09210-580, Brazil}

\begin{abstract}
New models of electrically charged objects as solutions of the Einstein-Maxwell field equations are obtained and studied in the present work. Static and rotating models are built. In both cases, the full spacetime geometries are obtained by matching two spacetime regions, an interior region containing electrified matter and an exterior electrovacuum region. In the static case, the interior region contains a spherically symmetric distribution of matter constituted by a de Sitter-type perfect fluid with electric charge, whose energy density profile is given by a Tolman-like relation. The interior solution is smoothly matched with the exterior Reissner-Nordstr\"om electrovacuum solution, thus producing different kinds of objects, such as charged regular black holes and overcharged tension stars, that we analyze in some detail. We also investigate the connection between the present static solution and the regular black holes with a de Sitter core presented in the work by Lemos and Zanchin [Phys. Rev. D {\bf83}, 124005 (2011)].
We then employ the G\"urses-G\"ursey metric and apply the
Newman-Janis algorithm to construct a charged rotating interior geometry from the static interior solution. The resulting interior metric and the electromagnetic field are smoothly matched to the exterior Kerr-Newman electrovacuum solution, thus producing a regular interior for the exterior Kerr-Newman geometry. The transition from the interior region to the exterior region of the spacetime is smooth in the sense that no boundary shell is present. The main properties of the complete rotating solutions are analyzed in detail, showing that different kinds of rotating objects, such as charged rotating black holes and other charged rotating objects, also emerge in the rotating case. 
\end{abstract}


\maketitle

\section{Introduction} 
The study of compact objects in General Relativity (GR) is attracting more and more attention thanks to the advances achieved by the large projects for collecting observational data, which culminated in the image of the astrophysical black hole in the giant elliptical galaxy M87~\cite{EHT-p6}, as well as in the detection of gravitational waves~\cite{LIGO2016PhRvL,LIGO2016ApJL,LIGO2016PRDfirstresults}. Historically, black holes have been among the most important and intriguing predictions of GR. As a consequence, a large amount of work on such structures has been carried out over the years, culminating in the remarkable singularity theorems~\cite{Penrose1965, HawkingEllis}, which led the formation of black holes to be accepted as a robust prediction of GR. However, GR also allows a wide variety of other kind of compact structures, such as ultra-compact stars~\cite{Bonnor:1972wi, bonnor75, lemoskleberzanchin, Lemos:2006sj, lemoszanchin2008}, gravastars~\cite{mazul, motto, visser, chirenti}, quasiblack holes~\cite{Lue1999, Lue2000, Lemos:2003gx, Lemos2007, Lemos2010}, and regular black holes~\cite{Bardeen, Frolov, Dymnikova92, Ayon, Bronni01, Dymnikova04, Hayward06, Broni06, Zaslavskii2010, Lemos2011, Lemos16, Fan2016, Masa18, Simpson19}.

In particular, regular black holes were proposed at about six decades ago.  
Indeed, in 1968, Bardeen \cite{Bardeen} concretely implemented the idea of a black hole with horizons but without a singularity, i.e., the first regular black hole model. The matter-energy content of such a regular black hole was later identified as a source of magnetic monopoles in nonlinear electrodynamics \cite{Ayon}, producing a modification of the Reissner-Nordstr\"om (RN) metric, so that, near the center, the energy content tends towards a perfect fluid with a de Sitter-type equation of state, where the pressure equals the negative of the energy density (in units such that the speed of light is unity). As well known, such a de Sitter fluid violates the strong energy condition, so that the singularity theorems do not apply. 
Although there has been great development in the implementation and analysis of the properties of regular black holes, most of the subsequent models of regular black holes in GR are based, in some sense, on the Bardeen's proposal.

The regular black hole models mentioned so far in this work are restricted to static configurations. As for the rotating counterparts,  much less models are found in the literature. However, motivated by the Newman-Janis complex transformation~\cite{Newman65}, G\"urses and G\"ursey~\cite{Gurses} obtained a stationary and axisymmetric metric in the Boyer-Lindquist coordinates from a metric belonging to the Kerr-Schild class~\cite{Schild}, and therefore, found a path to construct a rotating counterpart of any static and spherically symmetric solution. In such an approach, the resulting G\"urses-G\"ursey metric has the same form of the Kerr metric with the mass constant $m$ replaced by a mass function $m(r)$, where $r$ is the Boyer-Lindquist radial coordinate. Naturally, such strategy may also be used to find simple models for  rotating regular black holes starting from static black hole solutions of the Einstein equations.
Thenceforth, different paths were taken in order to better understand the type of sources that generates the G\"urses-G\"ursey metric~\cite{Burinskii, Gondolo}, as well as to generalize~\cite{Drake, Bambi} and modify~\cite{Mustapha} the Newman-Janis algorithm, what has led to a large amount of work on rotating regular objects~\cite{Spallucci, Modesto, Tosh14, Saa, Dymn15, Tosh17, Torres, Gosh, Mazza, Franzin, Masa22, Maeda, Brustein, Dymn2023, Casadio:2023iqt, Gosh23}.

The study of rotating regular black holes is motivated also by the known issues of the Kerr geometry~\cite{Kerr67}. Although it is accepted as the most realistic description of the exterior region of a black hole (outside the event horizon) with angular momentum, it presents some complications in the interior region of the black hole (inside the event horizon), such as a ring singularity, causality violation and closed timelike curves~\cite{Neil}. Hence, a possible route to avoiding such issues is replacing the problematic interior region with a regular matter source as done for static regular black holes. However, it is worth mentioning that the G\"urses-G\"ursey metric, on which most of the rotating regular black holes are built, depending on the form of the mass function $m(r)$, may present a conical or a curvature singularity located at $r=0$ and $\theta = \pi/2$ in Boyer-Lindquist coordinates. In fact, as discussed in \cite{Maeda, Ramon}, for regular mass functions that, close to $r=0$ is of the form $m(r) = m_0 r^{3 + \epsilon}$, with $\epsilon \ge 0$, presents a ring-like conical singularity when extended to the regions $r<0$. 
On the other hand, if the extension to the region $r < 0$ is
not performed, Torres~\cite{Ramon23} has shown that even the conical singularity at ring is absent and the spacetime is regular. Moreover, for $r m(r) \ge 0$, the G\"urses-G\"ursey metric is stably causal, which means that no closed causal curves appear even under small perturbations~\cite{Maeda}. Finally, it is noteworthy that the extension beyond the disk, through the region $r < 0$, is still a highly debated topic~\cite{Maeda, Ramon, Ramon23, Zhou23}.

Motivated by all the valuable contributions mentioned above, one of the main goals of the present work is constructing charged regular black hole and charged star solutions, both in the static and rotating cases. Our work is based on two main assumptions that we shall discuss in more detail next.

The first assumption comes from the objective of trying to avoid any singularity in the interior region of the matter source, in order to describe compact regular objects. As discussed earlier, the formation of singularities can be avoided in de Sitter-like spacetimes, which implies that the energy density and pressure of the fluid obey the relation $p = - \rho_m$, where $p$ is the isotropic pressure and $\rho_m$ is the matter energy density of the fluid. This assumption was also considered in the work by Lemos and Zanchin (hereafter L\&Z) \cite{Lemos2011} and it is widely used in the literature. 

The second assumption involves the energy density and is inspired by the Tolman VII solution. By performing a systematic study of the Einstein field equations applied to a static and spherically symmetric spacetime filled with a perfect fluid source, Tolman~\cite{Tolman39} obtained eight solutions in which three were already known at the time (the Einstein universe, the Schwarzschild-de Sitter solution, and the Schwarzschild star model), while the other five solutions were not known at the time. In particular, the Tolman VII solution appears to have some physical relevance in the literature, once it is an exact solution to the Einstein field equations that exhibits a quadratic falloff in the energy-density profile, i.e., the energy density is such that $8 \pi \rho_m(r) = 3R^{-2}\left(1 - {r^2}b^{-2}\right)$, where $r$ is the areal radius, and $R$ and $b$ are arbitrary constant parameters (see also Ref.~\cite{Raghoo}). Given that here we are going to consider charged perfect fluids, besides the energy density stemming from the matter distribution itself, the total energy density of the system includes the electromagnetic energy density carried by the electric field, the charged Tolman-like density relation we propose here is given by 
$8 \pi \rho_m(r) + q^2(r)\,r^{-4} = 3R^{-2}\left(1 - {r^2}b^{-2}\right)$, where $q(r)$ is the total electric charge inside the radius $r$. This profile is an electrified version of the relation postulated by Tolman to get his type VII solution. 
It is worth mentioning that, to the best of our knowledge, such a charged Tolman-like density profile has not been considered in the literature yet.
Let us also notice that, in the limit $b \to \infty$, we recover the assumption made by L\&Z \cite{Lemos2011} and first used by Cooperstock and de la Cruz~\cite{Cooperstock} and by Florides~\cite{Florides}. The resulting profile is an electrified version of the Schwarzschild assumption of constant energy density, and used to find the first interior solution for a compact object in general relativity, i.e., the Schwarzschild star (when glued with the Schwarzschild vacuum exterior).

After the two mentioned assumptions, the first step of this work is to generalize the solution obtained by L\&Z \cite{Lemos2011} that describes charged static regular black holes, as well as overcharged tension stars, recovering such a solution in the limit $b \to \infty$. By matching the interior solution to the RN exterior spacetime, we also find solutions that describe different kinds of regular black holes, as well as overcharged tension stars.

Beyond that, the other main goal of this work is to construct a charged rotating regular interior solution that can be matched to the exterior Kerr-Newman metric, once an interior regular continuation of the Kerr-Newman spacetime~\cite{newman1965}, with some "non too exotic" matter model, has not been achieved satisfactorily yet. By "non too exotic" matter we mean matter that "can be bought in the shops", paraphrasing H. Bondi~\cite{Zadeh}, for whom infinitely charged or infinitely massive matter, or matter moving at superluminal speed with respect to infinity are considered too exotic~\cite{Israel70, Frankowski, Hamity}. As an example of too exotic mater we mention the work by  Israel in~\cite{Israel70}, that found a disk-like interior solution composed of a material having negative energy density and rotating with superluminal velocity. The disk is the region spanned by $r = 0$ and $0 < \theta < \pi/2$ in B\"oyer-Lindquist coordinates, inside the singular ring of the Kerr and Kerr-Newman spacetimes defined by $r = 0$ and $\theta = \pi/2$. 
As summarized by Dymnikova~\cite{Dym}, the interior models for the Kerr and Kerr-Newman exterior solutions can be divided into disk-like~\cite{Israel70, Frankowski, Hamity}, shell-like~\cite{Lopez, Gron}, bag-like~\cite{Burinskii06}, and string-like~\cite{Israel77, Burinskii10} geometries.  See also Refs.~\cite{Hernandez-Pastora:2017fmg, Basso:2024hye} for more recent works on the subject.

Moreover, it is worth mentioning here that several solutions describing charged rotating regular black holes are obtained in the context of non-linear electrodynamics coupled to GR~\cite{Dymn15, Tosh17, Dymn2023}, with all of these dealing with continuous fields throughout the entire spacetime without any junction. In fact, most of the regular black hole solutions mentioned above have no definite boundary, see e.g.~\cite{Mustapha, Spallucci, Modesto, Tosh14, Saa, Torres, Gosh, Mazza, Franzin, Gosh23, Maeda}. Besides, several of these works do not investigate in detail the electromagnetic fields generated by the rotating charge distribution, particularly in those solutions obtained by applying the Newman-Janis algorithm.  
In contrast, the focus of our work lies in providing the derivation of all pertinent quantities within the interior region and properly match them with the exterior Kerr-Newman solution.
Aiming to build an interior rotating regular geometry for the Kerr-Newman exterior solution, in this work we follow the G\"urses-G\"ursey approach by starting with a well-defined charged non-rotating regular solution of the Einstein-Maxwell equations, which, as mentioned above, results from the electrified version of the Tolman VII energy-density profile. By matching the interior rotating solution with the Kerr-Newman exterior spacetime, we find all sorts of objects such as regular nonextremal black holes with a space, light and timelike boundary, regular extremal black holes with timelike  boundary and charged regular stars with timelike boundary. 

The present work is organized as follows. In order to set notation, in Sec.~\ref{sec:basequ} we present the fundamental equations for Einstein-Maxwell systems. In Sec.~\ref{sec:sco} we construct a charged static regular solution of the Einstein-Maxwell equations, which is smoothly matched together with
the exterior Reissner-Nordstr\"om solution, and analyze in detail the properties of the different kinds of objects that the complete solution gives rise. In Sec.~\ref{sec:rco}, by using the Newman-Janis procedure and the G\"urses-G\"ursey metric, we construct a charged rotating regular metric, which is smoothly matched together with the exterior Kerr-Newman metric, and analyze the properties of the different kinds of objects that the complete solution may describe. The possible interior electromagnetic fields that match appropriately the exterior Kerr-Newman fields are also analyzed in Sec.~\ref{sec:rco}.  Section \ref{sec:conc} is dedicated to final comments that summarize the main findings of the work and discuss aspects of the presented models of compact objects that deserve further investigation. Appendices~\ref{sec:appendixA} and \ref{sec:appendixEMT} contain additional material and general equations that are needed for the results shown in the main text.

\section{Fundamental equations}
\label{sec:basequ}

The compact objects considered in the present study are described by solutions of the Einstein-Maxwell field equations with electrically charged matter, i.e.,
\begin{align}
    & G_{\mu \nu} =  8\pi T_{\mu \nu}= 8\pi\left( E_{\mu \nu} + M_{\mu \nu}\right),\label{eq:Einst}\\
    &\nabla_{\nu} F^{\mu \nu} = 4 \pi J^{\mu}, \label{eq:Max1}\\
    & \nabla_{\alpha} F_{\mu \nu} + \nabla_{\nu} F_{\alpha \mu} + \nabla_{\mu} F_{\nu \alpha} = 0, \label{eq:Max2} 
\end{align}
where Greek indices range from $0$ to $3$.

The left-hand side of Eq.~\eqref{eq:Einst} is the Einstein tensor $G_{\mu \nu}= R_{\mu \nu}- \frac{1}{2}g_{\mu \nu} \mathcal R$, with $R_{\mu\nu}$ being the Ricci tensor, $g_{\mu \nu}$ being the metric tensor, and $\mathcal R$ being the Ricci scalar.  The right-hand side of Eq.~\eqref{eq:Einst} contains the energy-momentum tensor $T_{\mu \nu}$, which can be decomposed into two parts, $T_{\mu \nu} = E_{\mu \nu} + M_{\mu \nu}$, with the first part arising from the electromagnetic field, and the second part coming from the matter itself.

The Maxwell equations \eqref{eq:Max1}  and \eqref{eq:Max2} bear the Faraday-Maxwell electromagnetic tensor $F^{\mu\nu}$, which may be written in terms of a gauge vector potential $\mathcal{A}_{\mu}$ as $F_{\mu \nu} =  \nabla_{\mu} \mathcal{A}_{\nu} - \nabla_{\nu} \mathcal{A}_{\mu}$, with $\nabla_{\mu}$ standing for the covariant derivative compatible with the four-dimensional Lorentzian metric $g_{\mu\nu}$. Moreover, $J^\mu$ is the electromagnetic current density.

The electromagnetic energy-momentum tensor $E_{\mu \nu} $ is given in terms of $F_{\mu\nu}$ by
\begin{align} 
    E_{\mu \nu} = \frac{1}{4\pi}\left( F_{\mu \alpha} F_{\nu}^{\ \alpha} - \frac{1}{4}g_{\mu \nu} F_{\alpha \beta}F^{\alpha \beta}\right)\!.
\end{align}

The matter energy-momentum tensor $M_{\mu \nu}$ is given by a perfect fluid for static distributions, or by a non-isotropic fluid in the cases with rotation. 
 
\section{Charged static objects}
\label{sec:sco}
\subsection{Static and spherically symmetric systems: basic equations}
Here we restrict the analysis to static and spherically symmetric objects, where the charged fluid distribution is bounded by a spherical surface $\mathcal{B}_r$ defined by $r = r_0=\,$constant. 
As a consequence, the spacetime in both the interior and exterior regions of the fluid is static and spherically symmetric, which implies that the exterior region, in which $r \ge r_0$, is described by the electrovacuum Reissner-Nordstr\"om (RN) solution. In the interior region, the metric is conveniently written as
\begin{align}
ds^2 = - B_-(r) dt^2 + A_-(r) dr^2 + r^2d\Omega^2, 
\label{eq:intmetri}
\end{align}
where $(t, r, \theta, \varphi)$ are Schwarzschild-like coordinates, with the metric potentials $B_-(r)$ and $A_-(r)$ depending radial coordinate $r$ alone, and $d \Omega^2 = d \theta^2 + \sin^2 \theta d \varphi^2$ is the line element over the unit sphere. 

In turn, the electromagnetic gauge potential assumes the form
\begin{align}
\mathcal{A}_{-\mu} = -\phi_-(r) \delta_{\mu}^{\ t}, \label{eq:gaupot}    
\end{align}
where $\phi_-(r)$ is the scalar electric potential $\delta^{\mu}_{\ \nu}$ is the Kronecker delta.

The matter is assumed to be described by a perfect fluid, whose energy-momentum tensor is given by
\begin{align}
    M_{\mu \nu} = \big(\rho_m + p\big)u_{\mu}u_{\nu} + p g_{\mu \nu},
\end{align}
where $\rho_m$ is the energy density,  $p$ is the fluid pressure,  and $u^{\mu}$ is the fluid four-velocity satisfying $u^{\mu}u_{\mu} = -1$. 

In addition, we assume a convective current density $J^{\mu}_-$,
\begin{equation} 
J_{-}^{\mu} = \rho_e u^{\mu}, \label{eq:current}
\end{equation}
with $\rho_e$ standing for the electric charge density. 

The assumptions of staticity and spherical symmetry also implies that the fluid quantities $\rho_m$, $p$, and $\rho_e$ are also functions of the radial coordinate $r$ only. Moreover, in the regions where the Killing vector field $\xi^{\mu} = \delta^{\mu}_{\ t}$ is timelike, the velocity of the fluid $u^{\mu}$ is given by
\begin{align}
    u^{\mu} = B_{-}^{-1/2}(r) \delta^{\mu}_{\ t}. \label{eq:fourvel}
\end{align}

With the above assumptions, the $tt$ and $rr$ components of the Einstein field equations \eqref{eq:Einst} yield the following relations
\begin{align}
    & \frac{B'_-(r)}{B_-(r)} + \frac{A'_-(r)}{A_-(r)} = 8 \pi r A_-(r)\left[ \rho_m(r) + p(r) \right], \label{eq:tt} \\
    & \left(\frac{r}{A_-(r)}\right)' = 1 - 8 \pi r^2\left( \rho_m(r) + \frac{q^2_-(r)}{8 \pi r^4}\right), \label{eq:rr}
\end{align}
where the prime denotes differentiation with respect to the radial  coordinate $r$. 

The function $q_-(r)$ appearing in the last equation above is obtained in terms of the electric potential $\phi_-(r)$ from the $t$ component of the Maxwell field equations~\eqref{eq:Max1}. It gives
\begin{align}
   q_-(r) = - \frac{r^2 \phi'_-(r)}{\sqrt{A_-(r) B_-(r)}}= 4 \pi\! \!\int_0^r\!\! \rho_e(r) \sqrt{A_-(r)} r^2 dr, \label{eq:tmax}
\end{align}
where an integration constant has been set to zero. This relation means that $q_-(r)$ is the total electric charge inside the sphere of radius $r$. The second set of the Maxwell field equations~\eqref{eq:Max2} is trivially satisfied. 

Finally, the energy-momentum conservation equation $\nabla_{\nu} T^{\mu \nu} = 0$, together with Eq.~\eqref{eq:tmax}, implies the relation
\begin{align}
    2 p'(r) + \frac{B'_-(r)}{B_-(r)}\left[\rho_m(r) + p(r) \right] +  \frac{2 \rho_e(r) \phi'_-(r)}{\sqrt{B_-(r)}} = 0.\label{eq:enercon}
\end{align}

In summary, we have three independent differential equations, i.e., Eqs.~\eqref{eq:tt}, \eqref{eq:rr}, and \eqref{eq:enercon}, and five unknown quantities, namely, $A_-(r)$, $B_-(r)$, $\rho_m(r)$, $p(r)$, and $q_-(r)$. Hence, in order to close the system of equations, two further relations amongst these quantities must be supplied.

\subsection{Two additional assumptions}

The first additional assumption comes from the objective of avoiding any singularity in the interior region of the fluid,  in order to describe compact regular objects. As discussed earlier, the formation of singularities can be avoided by postulating negative fluid pressures. Here we assume a de Sitter-type equation of state, so that the energy density and pressure of the fluid obey the relation
\begin{align}
    \rho_m(r) + p(r) = 0. \label{eq:desit}
\end{align}

The second additional assumption is inspired by the Tolman VII solution \cite{Tolman39}. 
Such a solution appears to have some physical relevance in the literature, once it represents an exact solution to the Einstein field equations that exhibits a quadratic falloff in the energy density profile. Taking into account that, in the present case, we are considering charged perfect fluids, and that the total energy density includes the matter and also the electromagnetic energy density, we consider the following Tolman-like relation,
\begin{align}
8 \pi \rho_m(r) + \frac{q_-^2(r)}{r^4} = \frac{3}{R^2}\left(1 - \frac{r^2}{b^2}\right), \ \text{for} \ r \le r_0, \label{eq:tolden}    
\end{align}
where $R$ and $b$ are constant parameters introduced by Tolman \cite{Tolman39}, and that here are to be determined by the junction conditions. With our choice of units, both parameters $R$ and $b$ carry dimensions of length. The physical interpretation of $R$ is similar to the Schwarzschild interior solution, and as also considered in the case of charged matter interior solutions by Cooperstok and de La Cruz~\cite{Cooperstock},  and by Florides~\cite{Florides} (see also Ref.~\cite{Lemos2011}). In these models, parameter $R$ is such that $R^{-2}$ is proportional to the total energy density and, as the factor $3R^{-2}$ plays the role of the cosmological constant in the Einstein-Maxwell field equations, parameter $R$ has been interpreted as the de Sitter radius. In the present case, $R^{-2}$ is proportional to the matter energy density just at the center ($r=0$), and it is also interpreted as a de Sitter radius just as in the uncharged case.  In turn, parameter $b$ is the radius where the effective energy density vanishes and then, as we shall see below, it is related to the radius of the boundary surface of the matter distribution $r_0$.

\subsection{The electrically charged static spherical solution}
\label{sec:exsol}

\subsubsection{Interior solution} \label{sec:stinterior}

The first assumption given by Eq.~\eqref{eq:desit}, together with Eq.~\eqref{eq:tt}, implies the relation
\begin{align}
    B_-(r) = A^{-1}_-(r).
\end{align}
The second assumption given by the charged Tolman-like density in Eq.~\eqref{eq:tolden}, together with Eq.~\eqref{eq:rr}, give us $ B_-(r) = 1/A_-(r) = 1 + {m_0}/{r} -  {r^2}/{R^2} + {3\, r^4}/\left(5b^2 R^2\right)$, where $m_0$ is an integration constant. As well known, the term $m_0/r$ leads to a curvature singularity at $r\to 0$ and, since we are interested just in regular solutions, we put $m_0$ to zero. Hence, the metric potentials are given by
\begin{align}
    B_-(r) = A^{-1}_-(r) = 1 - \frac{r^2}{R^2} + \frac{3}{5}\frac{r^4}{b^2 R^2}.
    \label{eq:tolpot} 
\end{align}
As it is seen from Eq.~\eqref{eq:tolpot}, the present model, resulting from the hypotheses \eqref{eq:desit} and \eqref{eq:tolden}, furnishes the same metric potential of the type VII solution by Tolman \cite{Tolman39}. Such a metric potential has been employed as the seed to generate uncharged regular black holes in Ref.~\cite{Masa22}.

Now, we move forward in order to determine the fluid quantities. The assumption \eqref{eq:desit}, together with  the Maxwell equation \eqref{eq:tmax} and the static equilibrium equation \eqref{eq:enercon}, implies that
\begin{align}
    p'(r) =  \frac{1}{8\pi r^4} \frac{d}{dr}q^2_-(r). \label{eq:gradpre}
\end{align}
This relation, together with the two additional assumptions made in this work, namely, Eqs.~\eqref{eq:desit} and \eqref{eq:tolden}, allows us to find the electric charge function, 
\begin{equation}
    q_-(r) = \sqrt{\frac{3}{2}} \frac{r^3}{Rb},  \label{eq:q(r)}
\end{equation}
where, without loss of generality, we have chosen $q_-(r)$ to be positive. Notice that an integration constant results zero as a consequence of the field equations. Such a condition avoids singularities at $r=0$ in the electromagnetic energy-density.

From relations \eqref{eq:desit}, \eqref{eq:tolden}, and \eqref{eq:q(r)} it follows 
\begin{align}
& 8\pi\rho_m(r) = - 8\pi p(r) = \frac{3}{R^2}\left(1 - \frac{3r^2}{2b^2}\right). \label{eq:rhom}
\end{align}

It is worth noticing that the energy density of the fluid has a quadratic falloff from the center of the fluid distribution while the charge of the fluid distribution increases with the radius. Besides, as we shall see below, both the energy matter density and the pressure vanish at the surface of the fluid. As well, it is nice to see that, in the limit $b \to \infty$, we recover the interior solution found by Lemos and Zanchin~\cite{Lemos2011}, where the electric charge is spread over the surface of the fluid distribution and, therefore, there is no electric charge and electric field in the interior of fluid.

Given the relations~\eqref{eq:tolpot}, \eqref{eq:q(r)} and \eqref{eq:rhom}, all the important quantities of the problem may be determined.
For instance, the charge density profile $\rho_e(r)$ and the electric potential $\phi_-(r)$ are obtained from Eqs.~\eqref{eq:tmax} and \eqref{eq:q(r)}. They are given by
\begin{align}
    & \rho_e(r) = \rho_{e0}A^{-1/2}(r), \label{eq:rhoe} \\
    & \phi_-(r) =  -\sqrt{\frac{3}{8}} \frac{r^2}{R b} + \phi_0, \label{eq:phiint}
\end{align}
where $\rho_{e0}=3\sqrt{3/2}/(4 \pi Rb)$ and $\phi_0$ is an integration constant. Hence, the electric field $E_-(r)$ results in the form 
\begin{equation}
E_-(r) = \frac{q_-(r)}{r^2} = \sqrt{\frac{3}{2}} \frac{r}{R b}. \label{eq:EFint}
\end{equation}
Another important quantity is the mass $m_-(r)$ inside a sphere of radius $r$, which is given by
\begin{align}
    m_-(r) & \equiv 4 \pi \!\int_0^r\! \!\left(\rho_m(r) + \frac{q_-^2(r)}{8 \pi r^4} \right) r^2 dr + \frac{q_-^2(r)}{2 r},\label{eq:masssphe} \nonumber\\ 
    & = \frac{3}{2R^2}\left(\frac{r^3}{3} + \frac{3 r^5}{10 b^2} \right),
\end{align}
where $q_-(r)$ has been replaced from Eq.~\eqref{eq:q(r)}. Equation \eqref{eq:masssphe}
 allows us to write the metric potentials from Eq.~\eqref{eq:tolpot} as
 \begin{align}
     B_-(r) = A^{-1}_-(r) = 1 - \frac{2 m_-(r)}{r} + \frac{q^2_-(r)}{r^2}. \label{eq:stintAB}
 \end{align}

It is worth mentioning that the authors of Ref.~\cite{Tiwari}  obtained an equivalent interior solution  
by taking Eqs.~\eqref{eq:desit} and~\eqref{eq:rhoe} as initial assumptions. In contrast, our initial assumptions here are given by Eqs.~\eqref{eq:desit} and~\eqref{eq:tolden}. Therefore, we can see that both initial assumptions are equivalent.

\subsubsection{The exterior solution}
\label{sec:rnsol}
In the exterior region, for $r > r_0$, which corresponds to a electrovacuum region, the Einstein-Maxwell field equations for a static and spherically symmetric metric give us the well-known RN solution,
\begin{align}
ds_+^2 = - B_+(r) dt^2 + A_+(r) dr^2 + r^2d\Omega^2, 
\label{eq:rnmetric}
\end{align}
where
\begin{align}
    B_+(r) = A_+^{-1}(r) = 1 - \frac{2m}{r} + \frac{q^2}{r^2}, \label{eq:AB-RN}
\end{align}
with $m$ and $q$ being the mass and the electric charge parameters, respectively.

The solutions for the electric charge function, for the electric potential and for electric field in the exterior region are given by 
\begin{align}
    & q_{+}(r) \equiv q= {\rm constant},\label{eq:qrn}\\
    & E_{+}(r)= \frac{q}{r^2}, \label{eq:efrn}\\
  &  \phi_{+}(r)= \frac{q}{r}, \label{eq:phirn}
\end{align}
where the integration constant in the electric potential was set to zero.

\subsubsection{The matching conditions and the complete solution}
\label{sec:match}

Here we perform the smooth junction between the RN exterior solution and the
interior solution found in Sec.~\ref{sec:exsol} at the spherical boundary surface $\mathcal{B}_r$ defined by $r = r_0$. The interior region $\mathcal{M}_-$ corresponds to the spacetime domain for which $r < r_0$, while the exterior region $\mathcal{M}_+$ corresponds to the spacetime domain for which $r_0 < r < \infty$. We first deal with the metric junction conditions, by  employing the Darmois-Israel (DI) approach~\cite{Israel66},  and then we deal with the matching conditions of the electromagnetic field. 

Let $\xi^a = (\tau,\, \theta, \,\varphi)$ be the intrinsic coordinates of the boundary surface $\mathcal{B}_r$, and let $x^{\mu}_{\pm} = (t_\pm, \, r_\pm, \, \theta_\pm, \, \varphi_\pm)$ be the coordinates of regions $\mathcal{M}_-$ and $\mathcal{M_+}$, respectively. In fact, in view of the spherical symmetry of the geometry and of the boundary surface, the angular coordinates of the two spacetime regions may be identified, i.e., $\theta_-=\theta_+=\theta$, and $\varphi_-=\varphi_+=\varphi$. Additionally, without loss of generality, we may also identify the radial coordinates $r_-=r_+=r$. In turn, since we are dealing with smooth boundary conditions, the timelike coordinates of the two regions may also be identified, i.e., $t_-=t_+=t$ and, as a consequence, the intrinsic time coordinate on the boundary surface may be chosen as $\tau = t$. Therefore, from now on we drop the $\pm$ indexes. 
The DI approach deals with the first and the second fundamental form on $\mathcal{B}_r$, $h_{ab}$ and $K_{ab}$, respectively. Such geometric objects are defined in terms of the quantities in the four-dimensional objects by the well-known relations $h_{ab} = \varepsilon_a^{\ \mu} \varepsilon_b^{\ \nu}g_{\mu \nu}$, where $\varepsilon_{a}^{\ \mu} = \partial x^{\mu} / \partial \xi^a$, and $K_{ab} = \varepsilon_a^{\ \mu} \varepsilon_b^{\ \nu} \nabla_{\nu} n_{\mu}$, with $n_{\mu} = \sqrt{g_{rr}}\, \delta_{\mu}^{\ r}$ being the unit normal vector to the surface $\mathcal{B}_r$, and $\nabla_{\nu}$ stands for the covariant derivative compatible with the Lorentzian metric.

The assumption that the transition between the two spacetime regions is smooth implies that the first fundamental form on $\mathcal{B}_r$, $h_{ab}$ is continuous across the boundary, i.e., $[h_{ab}] = 0$. As usual the brackets $[...]$ indicate the difference of a given quantity across the boundary surface $\mathcal{B}_r$, i.e., $[Q] \equiv Q\big|^+_{\mathcal{B}_r} - Q\big|^-_{\mathcal{B}_r}$, with $Q\big|^{\pm}_{\mathcal{B}_r}$ representing any given quantity $Q$ evaluated on $\mathcal{B}_r$ from the point of view of the regions $\mathcal{M}_{\pm}$, respectively. The smooth transition across the boundary also requires that the second fundamental form, i.e., the extrinsic curvature $K_{a b}$ of $\mathcal{B}_r$, be a continuous function at the boundary, i.e., $[K_{ab}] = 0$. Therefore, by applying the smooth matching conditions, we find
\begin{align}
& \frac{r_0^2}{R^2}\left(2 - \frac{9 r_0^2}{5 b^2}\right) = \frac{m}{r_0}, \label{eq:match1} \\  
& \frac{3 r_0^2}{R^2}\left(1 - \frac{r_0^2}{b^2}\right) = \frac{q^2}{r_0^2}. \label{eq:match2}
\end{align}
It is straightforward to observe that, in the limit $b \to \infty$, we recover the L\&Z \cite{Lemos2011} matching conditions, namely, $Rq = \sqrt{3} r_0^2$ and $R^2 m = 2 r_0^3$.

In addition, when in the presence of electromagnetic fields in $\mathcal{M}_{\pm}$, such fields are also required to be matched together at the surface $\mathcal{B}_r$. The relevant electromagnetic quantities on $\mathcal{B}_r$ are the gauge potential $A_a = \varepsilon_{a}^{\ \mu}A_{\mu}$ and the field strengths $F_{ab} =  \varepsilon_a^{\ \mu} \varepsilon_b^{\ \nu} F_{\mu \nu}$ and $F_{an} = \varepsilon_a^{\ \mu} n^{\nu} F_{\mu \nu}$.
The smooth matching conditions for the electromagnetic fields are given by $[A_a] = 0$, $[F_{ab}] = 0$, and $[F_{an}] = 0$. In the present case,  the necessary smooth boundary condition $[A_a] = 0$ implies that $\phi_-(r_0) = \phi_+(r_0)$, where $\phi_\pm(r_0)$ stand for the electric potential in the corresponding region $\mathcal{M}_\pm$ evaluated at $r_- = r_0=r_+$.
By using Eqs.~\eqref{eq:phiint} and \eqref{eq:phirn} we find
\begin{equation}
    \phi_0 = \sqrt{\frac{3}{8}}\frac{r_0^2}{R\,b} + \frac{q}{r_0}. \label{eq:phi0}
\end{equation}
Furthermore, it is straightforward to see that the conditions $[F_{ab}] = 0$ are identically satisfied, while the conditions $[F_{an}] = 0$ implies that $E_+(r_0) = E_-(r_0)$. Hence, by using 
relations \eqref{eq:EFint} and \eqref{eq:efrn}, it follows
\begin{align}
    q= q_-(r_0)= \sqrt{\frac{2}{3}}\frac{r_0^3}{Rb}. \label{eq:qr0}
\end{align}

It is worth mentioning here that our work differs from the work by Lemos and Zanchin \cite{Lemos2011} once their model presents a shell of charge at the matching surface, and therefore the junction condition of the electromagnetic field is not smooth. Hence, as we shall discuss further ahead, the limit $b \to \infty$ of the present solutions requires a different junction condition for the electromagnetic field, in order to recover the solution of Ref.~\cite{Lemos2011}.

The smooth matching condition then yields three constraints, namely, Eqs.~\eqref{eq:match1}, \eqref{eq:match2}, and \eqref{eq:qr0}, for the five parameters of the model, namely $b,\, R,\,m,$ $q$, and $r_0$, so resulting that the present model has just two free parameters. Such two parameters may be chosen to be $r_0$ and $R$, so that it results
\begin{align}
    & b= \sqrt{\frac{3}{2}} r_0, \label{eq:br0R}\\
    & m = \frac{4}{5}\frac{r_0^3}{R^2}, \label{mr0R}\\
    & q= \frac{r_0^2}{R}. \label{eq:qr0R}
\end{align}

Additionally, by using Eqs.~\eqref{eq:match1}, \eqref{eq:match2}, \eqref{eq:phi0}, and \eqref{eq:qr0}, together with the expressions for $q_-(r)$, $M_-(r)$, $\rho_m(r)$, $p(r)$, $\phi(r)$, and $E_-(r)$,  
it is possible to completely characterize the quantities at the boundary surface $\mathcal{B}_r$. Namely, at that surface we have the relations
\begin{align}
& q_-(r_0) = q=\frac{r_0^2}{R}, \label{eq:qrR} \\ 
&  m_-(r_0) = m=\frac{4 r_0^3}{5 R^2} = \frac{4 q^2}{5 r_0}, \label{eq:mr0Rq} \\ 
& \rho_m(r_0) = p(r_0) = 0, \label{eq:rhomro}\\
& \phi_-(r_0) = \frac{q}{r_0}, \qquad E_-(r_0)= \frac{q}{r_0^2}. \label{eq:phiro}
\end{align}

Moreover, since $\phi_+(r_0) = q(r_0)/r_0 = q/r_0$, it follows from Eqs.~\eqref{eq:phi0}, \eqref{eq:br0R}, and~\eqref{eq:qrR} that the electric potential in the interior region  given by Eq.~\eqref{eq:phiint} may be cast into the form
\begin{align}
    \phi_-(r) = \frac{q}{2 r_0}\left(3 - \frac{r^2}{r_0^2}\right), \ 0\leq  r < r_0. \label{eq:intelecpot}
\end{align} 
As it can be easily seen, this potential has the same form of the electric potential in the interior region of a uniformly charged sphere with electric charge density given by $3q/4\pi r_0^3$ in flat spacetime physics.

In the present model, since the parameter $b$ is directly related to the radius of the boundary surface by Eq.~\eqref{eq:br0R}, it follows from relation \eqref{eq:rhom} that both the energy density and the pressure of the fluid vanish at the surface. On the other hand, in the solutions obtained by L\&Z \cite{Lemos2011}, the radius $r_0$ is not related to the parameter $b$. This difference comes from the use of distinct junction conditions for the electromagnetic field. Here, the junction condition is smooth and, therefore, from Eqs.~\eqref{eq:match2} and~\eqref{eq:qr0}, it is possible to obtain Eq.~\eqref{eq:br0R}. In contrast, the solution by L\&Z requires a shell of charge and, therefore, the junction condition for the electromagnetic field is not smooth. Another interesting fact to notice is that, for fixed $r_0$ and $R$, when compared to the present solution, the L\&Z solution requires a total electric charged larger by a factor of $\sqrt{3}$. Therefore, in order to describe compact regular objects, the solution obtained here requires less electric charge than the L\&Z solution.

\subsection{The complete static spherical solution and its main properties}
\label{sec:stclosed}

\subsubsection{The complete solution in closed form}

For the need of the next section, we write here the expressions of the most relevant quantities related to the static solution obtained above. 
The metric in each one of the spacetime regions $\mathcal{M}_-$ and $\mathcal{M}_+$ are then written in the form
\begin{equation}
ds_\pm^2 = -\left(1-\frac{2M_\pm(r)}{r}\right) dt^2 + \frac{dr^2}{1-\dfrac{2M_\pm(r)}{r}} + r^2d\Omega, \label{eq:stmetric}
\end{equation}
where 
\begin{align}
 &   M_-(r) = m_-(r) - \frac{q^2_-(r)}{2r} =  \frac{r^3}{2R^2}\left(1 - \frac {2}{5} \frac{r^2}{r_0^2} \right), \label{eq:m-(r)} \\
 &   M_+(r) = m - \frac{q^2}{2r}, \label{eq:m+(r)}
\end{align}
the $\pm$ signs go for $\mathcal{M}_\pm$, respectively.

Similarly, the electromagnetic gauge potential $\mathcal{A}_{\mu}$ is of the form 
\begin{equation}
    \mathcal{A}_{\pm\mu} = \phi_\pm(r)\,\delta_{\mu}^{\ t}\equiv \frac{Q_\pm(r)}{r}\delta_{\mu}^{\ t},  \label{eq:stgaugepot}
\end{equation}
where $\delta^{\mu}_{\ \nu}$ is the Kronecker delta, and $Q_\pm(r)$ is the charge function appearing in Eqs.~\eqref{eq:phiint} and \eqref{eq:qrn}, respectively. Namely,
 \begin{align}
      Q_-(r) & = \dfrac{q\, r}{2 r_0}\left( 3 - \dfrac{r^2}{r_0^2} \right),  \label{eq:Q-(r)}\\
      Q_+(r) &= q. \label{eq:Q+(r)}
 \end{align}

This gauge potential yields the following nontrivial components for the Faraday-Maxwell electromagnetic field tensor,  
\begin{align}
    & F_{\pm rt} = -F_{\pm tr}\equiv E_\pm(r), 
\end{align}
where
\begin{align}
& E_-(r) = \frac{r}{R r_0}=\frac{q\, r}{r_0^3},  \\
& E_+(r) = \frac{q}{r^2}.
\end{align}

The current density in each spacetime region is
\begin{align}
    & 4\pi\,J_-^\mu = \frac{3q}{r_0^3} \delta^\mu_t, \label{eq:stJ-}\\
    & 4\pi\,J_+^\mu = 0,  \label{eq:stJ+}
\end{align}
respectively.

\subsubsection{Curvature regularity}
 
The usual starting point in locating curvature singularities is by studying the behavior of the relevant curvature invariants, such as the Ricci scalar $\mathcal R = g_{\mu \nu}R^{\mu \nu}$, the Ricci squared $\mathcal{R}_2 = R_{\mu \nu} R^{\mu \nu}$, and the Kretschmann scalar $\mathcal{K} = R_{\mu \nu \lambda \sigma} R^{\mu \nu \lambda \sigma}$. In order to analyze the regularity of the solutions obtained in this work, we inspect the interior metric~\eqref{eq:intmetri} with the metric potentials $B_-(r)$ and $A_-(r)$ given by Eq.~\eqref{eq:tolpot}. Therefore, these curvature scalars are expressed by
\begin{align}
      & \mathcal R = \frac{12 }{R^{2}}\left(1 - \frac{r^2}{r_0^2}\right), \\
     & \mathcal{R}_2 = \frac{4}{R^{4}} \left(9 - \frac{18r^2}{r_0^2} + \frac{10r^4}{ r_0^4}\right), \\
     & \mathcal{K} =  \frac{8}{ R^4}  \left(3 - \frac{6 r^2}{r_0^2} + \frac{106}{25} \frac{r^4}{r_0^4}\right), 
\end{align}
which are clearly regular everywhere in the interior region $r < r_0$. In particular, at the origin $r = 0$, it follows
\begin{align}
    & \lim_{r \to 0 } \mathcal R = \frac{12}{R^2}, \\
    & \lim_{r \to 0 } \mathcal{R}_2 = \frac{36}{R^4}, \\
    & \lim_{r \to 0 } \mathcal{K} = \frac{24}{R^4}.
\end{align}
These limits are compatible with the results described in~\cite{Maeda}, which establish that, if the mass function $M_-(r) = m_-(r) - q_-^2(r)/2r$ may be expanded around $r = 0$ as $M_-(r) \approx c_0 r^{3 + \epsilon}$ with and $c_0 \neq 0$, then $r = 0$ is regular if, and only if,  $\epsilon \ge 0$.
To see this in the present case, it is enough to analyze behavior of the interior mass function $m_-(r)$ in the limit $r\to 0$.  
Equation \eqref{eq:m-(r)lz} gives $M_-(r) \approx r^3/2R^2$. 
As discussed in~\cite{Maeda}, for $\epsilon = 0$, we have $\lim_{r \to 0} R^{\mu \nu}_{\ \ \lambda \sigma} = 2 c_0\left(\delta^{\mu}_{\lambda}\delta^{\nu}_{\sigma} -\delta^{\mu}_{\sigma} \delta^{\nu}_{\lambda} \right)$ which implies that the spacetime is locally de Sitter once $c_0 > 0$. Moreover, as also shown by Maeda~\cite{Maeda}, for $\epsilon \ge 0 $, the tidal forces and Jacobi fields along an ingoing radial timelike geodesics also remain finite in the limit $r \to 0$. Therefore, there are no polynomial singularities at $r\to 0$.

\subsubsection{Energy conditions}
\label{sec:energcond}
In order to test the standard energy conditions, by following e.g. Hawking and Ellis~\cite{HawkingEllis}, let us first decompose the energy-momentum tensor in the local Lorentz frame, 
\begin{align}
    T_{a b} = e_{a}^{\ \mu} e_{b}^{\ \nu} T_{\mu \nu} = \text{diag}\left(\rho, \, p_1,\, p_2,\, p_3\right), \label{eq:emtst}
\end{align}
where $e_{a}^{\ \mu}, \ a  = 0,\,1,\,2,\, 3$, is a set of orthonormal basis vectors (a tetrad) that implements the local Lorentz frame and satisfies $\eta_{ab} = e_{a}^{\ \mu} e_{b}^{\ \nu} g_{\mu \nu}$, with $\eta_{ab}$ being the Minkowski metric. Once given the energy-momentum tensor in the form \eqref{eq:emtst}, we can state the standard energy-conditions as follows. The null energy condition (NEC) can be expressed as $\rho + p_i \ge 0$, for all $i = 1,\, 2,\, 3$. In turn, the weak energy condition  (WEC) requires  $\rho \ge 0$ in addition to the NEC. The dominant energy condition (DEC) requires $\rho - p_i \ge 0$ in addition to the WEC. The strong energy condition (SEC) requires $\rho + \sum_i p_i  \ge 0$ in addition to the DEC. While the DEC implies the WEC, both the WEC and the SEC include the NEC as a limiting case. Hence, all the standard energy conditions are violated if the NEC is violated.

The interior solution describes a static electrically charged perfect fluid whose energy-momentum tensor may be decomposed as $T_{\mu \nu} = E_{\mu \nu} + M_{\mu \nu}$, where the first part $E_{\mu \nu}$ arises from the electromagnetic field and the second part $M_{\mu \nu}$ comes from the matter itself. By defining the orthonormal tetrad as
\begin{align}
& e_0^{\ \mu} = B^{-1/2}_-(r) \delta^{\mu}_{\ t}, \qquad e_2^{\ \mu} = r^{-1} \delta^{\mu}_{\ \theta}, \\
 & e_1^{\ \mu} = A^{-1/2}_-(r) \delta^{\mu}_{\ r}, \qquad e_3^{\ \mu} = r^{-1} \sin^{-1} \theta \delta^{\mu}_{\ \varphi},
\end{align}
it follows that
\begin{align}
    & \rho(r) = \rho_m(r) + \frac{q^2_-(r)}{8 \pi r^4}, \label{eq:adens}\\
    & p_1(r) = p(r) - \frac{q^2_-(r)}{8 \pi r^4}, \label{eq:apres}\\
    & p_2(r) = p_3(r) = p(r) + \frac{q^2_-(r)}{8 \pi r^4}. \label{eq:tpres}
\end{align}

We first notice that the energy-momentum tensor of the charged fluid, including the contribution of the electromagnetic field, represents indeed an anisotropic fluid obeying the equation of state $\rho(r) + p_1(r) = 0$. Such a result follows from Eqs.~\eqref{eq:adens} and \eqref{eq:apres}, after using the assumption \eqref{eq:desit}.  Besides, from the expressions for $\rho(r)$ and for $p_i(r)$, $i=1,\,2,\,3,\,$ given above, together with the assumptions \eqref{eq:desit} and \eqref{eq:tolden}, it is easy to see that the NEC, the WEC, and the DEC are satisfied, while the SEC depends on the quantity
\begin{align}
    \rho + \sum_i p_i & = 2p(r) + \frac{q^2_-(r)}{4 \pi r^4} = \frac{1}{4 \pi R^2}\left(\frac{4 r^2}{r_0^2} - 1\right).
\end{align}
which is negative for $r^2 <r_0^2/4$. Therefore, considering non-negative values of $r$ only, the SEC is violated everywhere inside the surface $r = r_0/2$. This result is in agreement with the result due to Zaslavskii~\cite{Zaslavskii2010}, which has been extended by Maeda~\cite{Maeda}, and that may be stated as follows. If, close to $r=0$, the mass function may be expanded in the form $M_-(r) = c_0 r^{3 + \epsilon} + c_1 r^{3 + \epsilon + \beta}$, with $\epsilon \ge 0$, $\beta > 0$, $c_0 > 0$, and $c_1 < 0$, then the NEC, the WEC, and the DEC are satisfied, while the SEC is violated around $r = 0$.

\subsubsection{On the horizons and other properties of the solution}
\label{sec:sthorizons}

Let us now investigate the existence of horizons and other properties of the complete static spherical solution presented above.  For that, it is necessary to properly parameterize the relevant relations involved in the present analysis. 
As mentioned in Sec.~\ref{sec:match}, among the five parameters $b$, $R$, $r_0$, $q$, and $m$, the complete solution has only two free parameters. For the sake of comparison, we follow Lemos and Zanchin~\cite{Lemos2011} and express the parameters $b$, $R$, $r_0$, and $m$ in terms of $q$ and of a new parameter $\alpha$ given by $r_0= \alpha q$. With this, Eqs.~\eqref{eq:br0R}--\eqref{eq:qr0R} allow us to obtain the relations
\begin{align}
& r_{0} = \alpha q,\\
& R = \alpha^2 q,\\
& m = \frac{4}{5 \alpha}q.
\end{align}
Hence, from now on,  $\alpha$ and $q$ shall play the role of free parameters in the present model.

As it is well known, for static spacetimes, the existence of horizons is indicated by the presence of spacetime regions where the Killing vector field $\xi^{\mu} = \delta^{\mu}_{\ t}$ becomes lightlike, a condition that, in the present case, corresponds to the vanishing of the metric potential $B_{\pm}(r)$. For simplicity, we dub such a metric potential the horizon function. 

We first consider the possible horizons in the spacetime region $\mathcal{M}_-$, i.e., inside the matter distribution, where the horizon function $B_-(r)$ is given by Eq.~\eqref{eq:tolpot}. 
By imposing $B_{-}(r) = 0$, we find two possible non-negative roots given by
\begin{align}
r_{k-} = \sqrt{\frac{5r_0^2}{4}\left(1 +(-1)^k \sqrt{1 - \frac{8}{5} \alpha^2}\right)}, \label{eq:stinhoriz}  
\end{align}
where $k = 1$ indicates the smaller radius, while $k = 2$ indicates the largest radius,
and the negative roots were disregarded since the radial coordinate is assumed to be positive. 

\begin{figure}[t!]
    \centering
        \includegraphics[width=8.5cm]{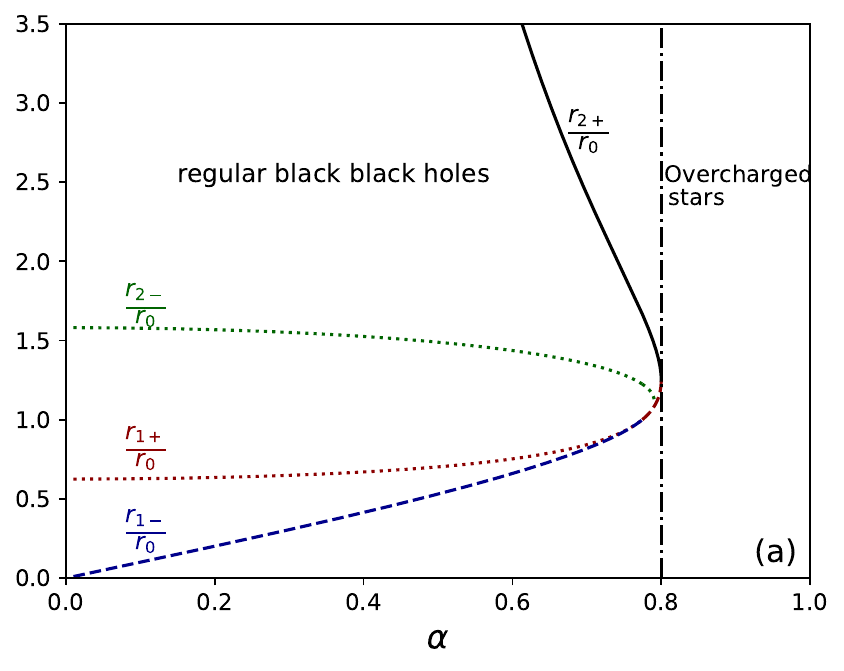}
        \includegraphics[width=8.3cm]{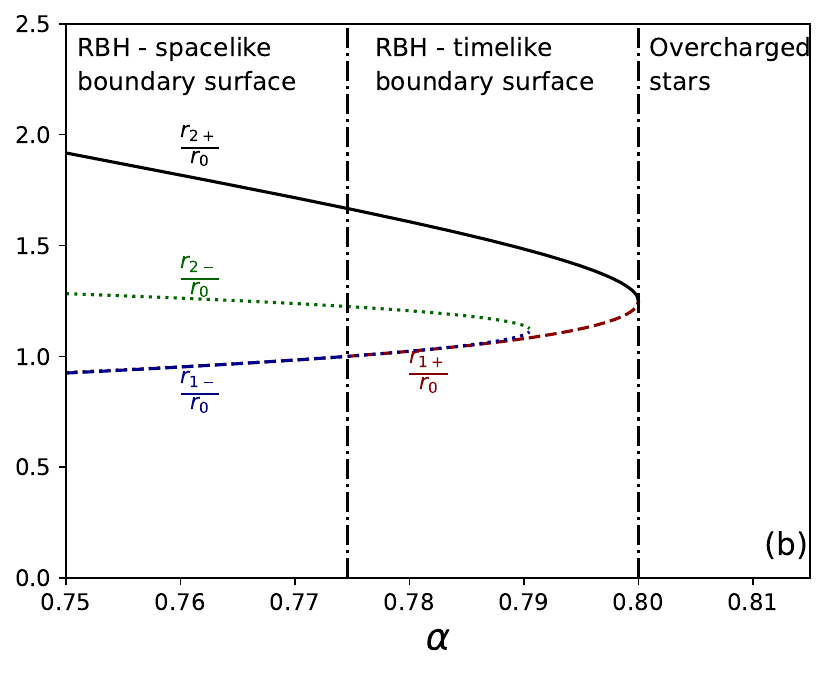}
    \caption{Panel (a), top: A plot of the normalized radii $r_{1\pm}/r_0$ and $r_{2\pm}/r_0$ as a function of $\alpha=r_0/q$, with labels identifying each curve. The dashed line represents the true Cauchy horizon, while the solid line represents the true event horizon. The dotted lines represent the different radii out of their domain of validity. The vertical line separates the region of regular black holes (RBH) solutions from the region of overcharged stars solutions. \\
    Panel (b), bottom: The region of the top panel for $\alpha$ in the interval $0.75 < \alpha < 0.82$ zoomed in. The first vertical dashed-dotted line at $\alpha = \sqrt{3/5}$ separates the region of RBH with spacelike boundaries from the region of RBH with timelike boundaries. The second vertical dashed-dotted line at $\alpha = 4/5$ separates the region of RBH with timelike boundaries from the region of overcharged stars.}
    \label{fig:radii}
\end{figure}

As it can be seen from Eq.~\eqref{eq:stinhoriz}, for $\alpha > \sqrt{5/8}$, both roots $r_{k-}$ assume complex values and, therefore,  $B_-(r)$ has no horizons. On the other hand, for $0 < \alpha \le \sqrt{5/8}$, the two roots $r_{k-}$ are real valued numbers and may represent horizons. However, 
in order for the roots $r_{k-}$ being actual horizons the conditions $r_{k-} \leq r_0$ must be also satisfied. The condition $r_{2-} \leq r_0$ cannot be satisfied for any real value of $\alpha$ in the interval $0 < \alpha \le \sqrt{5/8}$.  In fact, as depicted in panel (a) of Fig~\ref{fig:radii}, the larger root $r_{2-}$ is such that $r_{2-}/ r_0> 1$ in the whole interval $0 < \alpha \le \sqrt{5/8}$, and hence the root $r_{2-}$ stands in a region of the spacetime where the interior solution is no longer valid. As a consequence, the spacetime region $\mathcal{M}_-$ possesses at most one horizon. As a matter of fact, the condition  $r_{1-} \leq r_0$ implies that $\alpha^2\leq 3/5$. Therefore, the spacetime may present a horizon in the region $\mathcal{M}_-$ just for $\alpha$ in the interval $0< \alpha < \sqrt{3/5}$.

For $\alpha = \sqrt{3/5}$, the smaller real root in \eqref{eq:stinhoriz} coincides with the boundary of the fluid (the matching surface $\mathcal{B}_r$), i.e., $r_{1-}=r_0$. This implies that, for $\alpha$ in the interval $\sqrt{3/5} < \alpha \le \sqrt{5/8}$, both zeros of $B_-(r)$ are such that $r_{k-} > r_0$, corresponding to a region of the spacetime where the interior solution is no longer valid. Therefore, whenever $\alpha > \sqrt{3/5}$, or equivalently for $r_0 < \sqrt{5/3}\,R$, the interior metric potential $B_-(r)$ presents no real roots in its domain of validity. In comparison with the L\&Z solution~\cite{Lemos2011}, their interior solution has no horizons for $r_0 < R$.

The possible horizons of the exterior spacetime region $\mathcal{M}_+$ are given by the roots of the RN metric coefficient $B_+(r)$ presented in Eq.~\eqref{eq:AB-RN}, which gives the well-known results $ r_{k+} = m + (-1)^k \sqrt{m^2 - q^2}$, with $k = 1,2$. In terms of the free parameters $\alpha$ and $q$, these may be written as
\begin{align}
  \label{eq:RNhor}
  r_{k+} = \frac{4 r_0}{5\alpha^2}\left(1 + (-1)^k \sqrt{1 - \frac{25}{16} \alpha^2}\right). 
\end{align}
Notice that booth roots $r_{k+}$ assume real non-negative values just for $\alpha$ in the interval $0\leq \alpha\leq 4/5$. Additionally, these roots must satisfy the conditions $r_{k+} \geq r_0$. The condition $r_{2+}\geq r_0$ is satisfied for all $\alpha$ in the interval $0< \alpha \leq 4/5$. Since $r_{2+}$ increases with $\alpha^{-1}$, it suffices to check such a condition for the largest value, $\alpha=4/5$, which gives $r_{2+}=5r_0/4> r_0$. 
The condition $r_{1+}\geq r_0$ furnishes $\alpha\geq \sqrt{3/5}$. 
Therefore, we reach the following conclusions.  \vskip .1cm
\noindent
{\it i}) For $\alpha$ in the interval $0 < \alpha \leq 4/5$, the complete solution shall present just two horizons, an event horizon and a Cauchy horizon.\vskip .1cm
\noindent
{\it ii}) For $\alpha$ in the interval $4/5 < \alpha < \infty$, the complete solution shall present no horizons. \vskip .1cm
\noindent
{\it iii}) 
The root $r_{2+}$, given in Eq.~\eqref{eq:RNhor}, shall be the event horizon for all different kinds of solutions with $\alpha$ in the interval $0 < \alpha \leq 4/5$. \vskip .1cm
\noindent
{\it iv}) The root $r_{1+}$, given in Eq.~\eqref{eq:RNhor}, shall be the Cauchy horizon for all different kinds of configurations with $\alpha$ in the interval $\sqrt{3/5} \leq \alpha \leq \sqrt{5/8}$.\vskip .1cm 
\noindent
{\it v}) The root $r_{1-}$, given in Eq.~\eqref{eq:stinhoriz}, shall be the Cauchy horizon for all different kinds of solutions with $\alpha$ in the interval $0 < \alpha \leq \sqrt{3/5}$. \vskip .1cm 
\noindent
{\it vi}) The root $r_{2-}$, given in Eq.~\eqref{eq:stinhoriz}, is larger than $r_0$ for all $\alpha$ and then it has no physical meaning. \vskip .1cm 

{\noindent \it vii}) The limit of zero charge with finite $r_0$ is obtained by taking $\alpha \to \infty$. Then, Eq.~\eqref{eq:qr0R} implies that either $r_0 \to  0$ or $R \to \infty$. In the case $r_0 \to 0$ the matter region shrinks to a point, while in the case $R \to \infty$ the energy density and the pressure vanish everywhere. Hence, excluding the singular solutions, both cases result in the Minkowski spacetime.

 
The properties just listed above are illustrated in panels (a) and (b) of Fig.~\ref{fig:radii}. More details on the different kinds of objects the present solution of the Einstein-Maxwell equations generates are given next.

\subsection{The different kinds of objects in the complete static solution}

\subsubsection{Regular nonextremal black holes with a spacelike boundary}

Considering $\alpha$ in the interval $0< \alpha < \sqrt{{3}/{5}}$, it follows that the roots $r_{1-}$ and  $r_{2+}$ assume real values and satisfy the inequalities $r_{1-} < r_0 < r_{2+}$.  Moreover, the mass to charge ratio of each one of the objects with $\alpha$ in such an interval is in the range $4/\sqrt{15} < m/q < \infty$. We recall from the last section that $r_{1-}$ is the smallest real root of the horizon function defined in region $\mathcal{M}_-$, given in Eq.~\eqref{eq:stinhoriz}, while  $r_{2+}$ is the largest real root of the  horizon function defined in region $\mathcal{M}_+$, given in Eq.~\eqref{eq:RNhor}. Therefore, $r_{1-}$ is the Cauchy horizon and is inside the matter distribution, while $r_{2+}$ is the (RN) event horizon of the complete solution. Since the radius of the boundary surface $r_0$ lies between the two true horizons, the corresponding configurations are regular nonextremal black holes with a spacelike boundary. These charged regular black hole solutions are similar to the uncharged solution analyzed by Frolov et al~\cite{Frolov90}, where the exterior Schwarzschild geometry is glued to the interior de Sitter geometry at a spacelike surface inside the horizon $r = 2m$, the main difference being that our model does not present a transition layer.
The configurations generated by $\alpha$ in the interval $0< \alpha < \sqrt{{3}/{5}}$ correspond to great part of the region labeled as "regular black holes region" in panel (a) of Fig.~\ref{fig:radii}. The related region extends from arbitrarily small $\alpha> 0$ to the first vertical dashed line that is shown just in panel (b) of Fig.~\ref{fig:radii}.

\subsubsection{Regular nonextremal black holes with a lightlike boundary at the Cauchy horizon}

In the particular case where $\alpha = \sqrt{3/5}$, the roots $r_{1-}$ and $r_{1+}$ coincide, $r_{1-}=r_{1+}$, and it also follows $r_{1-}=r_0=\sqrt{3/5}\, q$,  while $r_{2+} >r_0$, i.e., the radius of the boundary surface $r_0$ coincides with the Cauchy horizon and the matter fills the interior region up to $r=r_{1-}=r_{1+}$. The event horizon is at $r_{2+}$, which attains the value $r_{2+} = q/\alpha =  \sqrt{5/3}\,q=5r_0/3$. The mass to charge ratio of each one of the corresponding kind of objects is $m/q=4/\sqrt{15}\simeq 1.033$.
From Eq.~\eqref{eq:rhomro}, we see that both the matter energy density $\rho_m$ and the pressure of the fluid $p$ vanish at the boundary surface $r_0$. The same happens to the charge density $\rho_e(r)$ since, from Eq.~\eqref{eq:rhoe}, we have $\rho_e(r_0) = \rho_{e0}\sqrt{B(r_0)} = 0$. Therefore, this solution represents a regular nonextremal black hole with a lightlike boundary at the Cauchy horizon, 
with no matter nor electric charge at the boundary surface.
This situation corresponds to the first vertical dashed-dotted line shown in panel (b) of Fig.~\ref{fig:radii}.

It is worth mentioning that the lightlike matching surface is obtained here as the limiting case of configurations with timelike boundaries, where the radius of the timelike boundary gets arbitrarily close to the Cauchy horizon. However, more properly, it can also be obtained by using the smooth junction conditions for lightlike boundaries. 
The steps to follow in such an approach are the same as those presented in Appendix A of Ref.~\cite{Lemos2011}.

\subsubsection{Regular nonextremal black holes with a timelike boundary}

Another interesting class of compact objects are found for values of the parameter $\alpha$ in the interval $\sqrt{3/5}< \alpha < {4}/{5}$.
As commented in Sec.~\ref{sec:sthorizons}, whenever $\alpha > \sqrt{3/5}$, the interior metric potential $B_-(r)$ presents no real roots in its domain of validity. In fact, 
for $\alpha$ in the interval $\sqrt{3/5}< \alpha < {4}/{5}$, we have $r_{k-}> r_0$ for both $k=1$ and $k=2$, and so the horizons are given by the roots $r_{k+}$, whose values are both larger than the radius of the boundary surface, i.e., $r_0 < r_{1+} < r_{2+}$. In this region, the Cauchy horizon $r_{1+}$
grows with $\alpha$ and remains outside the radius of the matter boundary, while the event horizon $r_{2+}$ shrinks with $\alpha$ but remains larger than the Cauchy horizon. There is matter up to $r_0$, with both the matter energy density and the pressure vanishing at $r_0$, and then the two horizons stand outside the matter region. The gravitational mass $m$ is slightly larger than the electric charge and bounded by 
$1 < {m}/{q} < {4}/{\sqrt{15}}$. 
Hence, in this interval of $\alpha$, the configurations are regular nonextremal black holes with a timelike boundary inside the inner horizon. 
This situation corresponds to the region between the two vertical dashed-dotted lines shown in panel (b) of Fig.~\ref{fig:radii}.

\subsubsection{Regular extremal black holes with a timelike boundary }

In the particular case where $\alpha = 4/5$, the roots $r_{1+}$ and $r_{2+}$ define the Cauchy horizon and the event horizon, respectively. In this limiting case, the two horizons coincide and are given by $r_{1+} = r_{2+} = q = m$, and it also follows $r_0 = 4 r_{2+}/5$. Again, both the matter energy density $\rho_m$ and the pressure $p$ vanish at $r_0$, but differently from the other particular case with $\alpha=\sqrt{3/5}$, the charge density does not vanish at $r=r_0$. In the present case one has $r_0<r_{1+}=r_{2+}$, so that the degenerate horizon stands at a further coordinate distance from the boundary surface. Therefore, the corresponding configuration represents a regular extremal black hole with a timelike boundary inside the degenerate horizon. This situation corresponds to the vertical dashed-dotted line shown in panel (a) of Fig.~\ref{fig:radii}, which is the same as the second vertical dashed-dotted line shown in panel (b) of Fig.~\ref{fig:radii}.

\subsubsection{Regular overcharged stars with a timelike boundary}

For all values of the parameter $\alpha$ larger than $\alpha = 4/5$, it follows that $q > m$ meaning we are dealing with overcharged objects. 
Moreover, when $\alpha> 4/5$, Eq.~\eqref{eq:RNhor} implies that there are no horizons and so no black holes. Therefore,  overcharged stars, with no horizon and charge greater than mass, come into being. In fact, the mass to charge ratio of each one of these objects is in the range $0 < m/q < 1$.
This situation correspond to the region on the right of the last vertical dashed-dotted lines in panel (a) of Fig.~\ref{fig:radii} and in panel (b) of Fig.~\ref{fig:radii}.

\subsection{The limit to the Lemos and Zanchin electrically charged solution}
\label{sec:lemzan}

We analyze here how to take the proper limit $b \to \infty$ of the charged static solution presented in Sec.~\ref{sec:sco} to recover the L\&Z electrically charged solution \cite{Lemos2011}. 
That solution also describes charged regular black holes and overcharged tension stars with the RN solution outside a de Sitter core. 
Since the exterior solutions are the same in both models, and they do not depend explicitly on the parameter $b$, here we need to deal just with the interior solution and, afterwards, with the matching conditions. 

We start by taking the limit $b \to \infty$ in Eq.~\eqref{eq:tolden}. It implies  the inequality
\begin{align}
    8 \pi \rho_m(r) + \frac{q^2_-(r)}{r^4} = \frac{3}{R^2}, \quad \text{for} \ r \le r_0, \label{eq:elecschw}  
 \end{align}
which is exactly the initial assumption made in the work of Ref.~\cite{Lemos2011}.
The limit of other fluid quantities follows in a similar way. For instance, Eq.~\eqref{eq:gradpre} reduces to $q_-(r)=0$, and this result, together with Eq.~\eqref{eq:rhom}, gives 
\begin{align}
& \rho_m(r) = - p(r) = \frac{3}{8 \pi R^2}, \quad \text{for} \ r \le r_0, \label{eq:rhomlz}
\end{align}
exactly the result by L\&Z \cite{Lemos2011}.

The total mass of the L\&Z solution inside a surface of radius $r$ also may be obtained by taking the limit $b\to\infty $ of the mass function $m_-(r)$ of the present solution, as given in Eq.~\eqref{eq:masssphe}. The result is
\begin{align}
    m_-(r)  = \frac{r^3}{2R^2}. \label{eq:m-(r)lz}
\end{align}

Analogously, the metric potentials, which can be obtained by taking the limit $b \to \infty$ in Eq.~\eqref{eq:tolpot}, gives the expected result for the interior metric functions,
\begin{align}
       B_-(r) = A^{-1}_-(r) = 1 - \frac{r^2}{R^2},
    \label{eq:tolpotlz} 
\end{align}
what completes the interior metric solution of L\&Z.

The interior electric potential $\phi_-(r)$ and the electric field $E_-(r)$ of the L\&Z solution are also obtained from the present solution by taking the limit $b \to\infty$ in the corresponding expressions, Eqs.~\eqref{eq:phiint} and \eqref{eq:EFint}, which furnish the expected results for such quantities,  
\begin{align}  
 \phi_-(r)  = \phi_0, \qquad E_-(r)  = 0,   \label{eq:phiintlz}
 \end{align}
where $\phi_0$ is a constant.

Now we turn attention to the matching conditions. The smooth junction of the first and second fundamental forms may be obtained directly from Eqs.~\eqref{eq:match1} and \eqref{eq:match1}. By taking the limit of large $b$ in such relations, we get
$2r_0^3  = m\, R^2$ and $3 r_0^4  = q^2R^2$, 
which are exactly the L\&Z~\cite{Lemos2011} matching conditions.

On the other hand, the matching of the electromagnetic fields is more intricate. In fact, the correct junction of the L\&Z solution for the electric potential $\phi(r)$ can be obtained by taking the limit $b\to\infty$ in Eq.~\eqref{eq:phi0}, and then by taking into account that $\phi(r)$ is continuous across the boundary. Namely, 
\begin{equation} \label{eq:phipmlz}
   \phi(r) =\left\{ \begin{array}{ll}
    \dfrac{q}{r_0}, & \qquad {\rm for\;} 0\leq r\leq r_0,\\   
   \! \!\dfrac{q}{r}, & \qquad {\rm for\;} r\geq r_0. 
 \end{array} \right.
\end{equation}

However, the correct junction of the L\&Z solution for the electric field $E(r)$ does not follow just by taking the limit $b\to\infty$ in Eq.~\eqref{eq:qr0}, which would furnish a vanishing electric charge and a vanishing electric field strength throughout spacetime.
The jump of the electric charge function at the boundary surface, from $q_-(r)= 0$ to $q_+(r)=q=$ constant, implies that the electric field strength $E(r)$ also presents a jump at $r=r_0$, i.e., 
\begin{equation} \label{eq:matchElz}
E(r) =\left\{ \begin{array}{ll}
     0, & \qquad {\rm for\;} 0\leq r < r_0,\\   
     \dfrac{q}{r^2}, & \qquad {\rm for\;} r\geq r_0. 
 \end{array}\right.
\end{equation}
In our notation, this jump reads $ [F_{an}] = 4 \pi \sigma_e u_a$, with $\sigma_e$ and $u_a$ being the proper surface charge density and the proper four velocity of the shell, respectively. In fact, the proper charge density is given by 
 \begin{equation} 
 \sigma_e = \dfrac{q}{4\pi\, r_0^2}. \label{eq:stchd}
 \end{equation}
As it is seen, the only region of spacetime bearing some electric charge is the boundary surface, which is uniformly electrified.

\section{Charged rotating objects}
\label{sec:rco}

\subsection{The G\"urses-G\"ursey metric and the electromagnetic field}
\label{sec:ggmetric}

In the particular case where $A(r)\times B(r)=1$, a rotating version of the metric \eqref{eq:stmetric} may be generated by the G\"urses-G\"ursey approach \cite{Gurses}. 

The mentioned approach then furnishes the rotating metric, which, in the Boyer-Lindquist coordinate reads
\begin{align}
   ds_\pm^2  = & -\left(1 - \frac{2 r\, M_\pm(r) }{\Sigma} \right) dt^2 + \frac{\Sigma}{\Delta_\pm} dr^2 \nonumber \\ & + \Sigma d\theta^2 - \frac{4 r\, M_\pm(r)  a \sin^2 \theta}{\Sigma} dt\, d\varphi \label{eq:ggmetric} \\
    & + \left(r^2 + a^2 + \frac{2 r\, M_\pm(r) a^2 \sin^2 \theta}{\Sigma} \right) \sin^2\theta\, d\varphi^2, \nonumber
\end{align}
where $a$ is the rotation parameter that is taken to be the same in both spacetime regions, and $\Sigma$ and $\Delta_ \pm$ stand for 
\begin{align}
& \Sigma \equiv \Sigma(r,\theta)= r^2+ a^2\cos^2\theta, \label{eq:sigmaa}\\
   &\Delta_\pm\equiv \Delta_\pm (r) = r^2 + a^2 - 2r\, M_\pm(r), \label{eq:deltfun}
\end{align}
respectively. The mass function $M_\pm(r)$ stands for 
\begin{equation}
M_\pm(r) = m_\pm(r) - \frac{q_\pm^2(r)}{2r}, \label{eq:massM}
\end{equation}
where $m_\pm(r)$ indicates the gravitational (ADM) mass and $q_\pm(r)$ indicates the electric charge.

Similarly, the electromagnetic gauge potential and the field strength in the rotating metric \eqref{eq:ggmetric} may be obtained from the appropriate transformation of the static gauge potential $\mathcal{A}_{\pm\mu}$ \eqref{eq:stgaugepot}.
Then, to deal with the electromagnetic gauge potential, we follow closely the type-I complexification introduced by Bambi and Modesto in~\cite{Bambi}. In comparison to the work by Bambi and Modesto, we also have to take into account the electromagnetic field. For this, we need to write the electric potential $\phi(r)$ of the nonrotating solution as $\phi(r) = Q(r)/r$ and follow the same procedure as for the mass function, i.e., transforming the factor $1/r$ without changing the function $Q(r)$. Hence, we get
\begin{align}
    \mathcal{A}_{\pm\mu} = -\frac{Q_\pm(r)r}{\Sigma}\left(\delta_{\mu}^{\ t} - a \sin^2 \theta \delta_{\mu}^{\ \varphi}\right), \label{eq:rotpot} 
\end{align}
where the charge functions $Q_\pm(r)$ are given by Eqs.~\eqref{eq:Q+(r)} and \eqref{eq:Q-(r)}, respectively.
This gauge potential yields the following nontrivial components for the electromagnetic field tensor,  
\begin{align}
    & F_{\pm rt} = \frac{Q_\pm(r)\left(r^2 - a^2 \cos^2 \theta\right) }{\Sigma^2} - \frac{r\, Q_\pm'(r)}{\Sigma}, \nonumber \\
    & F_{\pm \theta t} = - \frac{Q_\pm(r)\, r\, a^2 \sin 2 \theta}{\Sigma^2}, \label{eq:electen} \\
    & F_{\pm r\varphi} = - a \sin^2 \theta F_{\pm rt}, \nonumber \\ 
    & F_{\pm\theta \varphi} = -\frac{\left(r^2+a^2\right)}{a} F_{\pm\theta t},  \nonumber
\end{align}
where the prime indicates the total derivative with respect to the coordinate $r$. 
It is worth mentioning that the use of the charge function $Q(r)$ in the expression for electromagnetic gauge potential~\eqref{eq:rotpot} and for the electromagnetic field tensor~\eqref{eq:electen}, instead of total electric charge inside a radius $r$ given by $q(r)$, is justified since it allows us to smoothly match together the interior and the exterior electromagnetic fields across the boundary, as we show in Sec.~\ref{sec:matchr}.

In Appendix~\ref{sec:appendixA} we show that the electromagnetic field tensor~\eqref{eq:electen} satisfies the Maxwell equations. Let us mention that, in their seminal work,  Newman {\it et al}.~\cite{newman1965} did not apply the same algorithm in order to obtain the electromagnetic strength tensor of the Kerr-Newman solution. The authors derived the rotating metric following the Newman-Janis algorithm, and then calculated the electromagnetic strength by direct integration of the Maxwell equations. As shown in Appendix~\ref{sec:appendixA}, the Faraday-Maxwell tensor \eqref{eq:electen} obeys the Maxwell equations, and the nature of the current density of the interior region $J_{-}^{\mu}$ is also discussed there.

For now let us determine the total charge $\mathfrak{q}_{-}(r)$ inside a spheroid of $r = \text{constant}$ in the interior region, i.e., for $r\leq r_0$. Given a surface $\varSigma_t$ defined by $t  = \text{constant}$, such a charge can be defined as
\begin{align}
    \mathfrak{q}_{-}(r) = -\int_{\varSigma_t}^r J_-^{\mu} \eta_{\mu} d \mathcal{V} = -\frac{1}{4\pi} \int_{\varSigma_t}^r \nabla_{\nu} F_-^{\mu \nu} \eta_{\mu} d \mathcal{V}, \label{eq:spherq}
\end{align}
where $\eta_{\mu}$ is a unitary time-like vector orthogonal to the surface $\varSigma_t$ and $d \mathcal{V} = \sqrt{h}\, dr\, d \theta\, d \varphi$, with $h$ being the determinant of the induced metric in $\varSigma_t$. The upper index $r$ in the integration sign indicates that the integration is not taken over the whole hyper-surface $\varSigma_t$, but just inside the spheroid $r=$ constant. 
A straightforward calculation shows that

\begin{align}
    \mathfrak{q}_{-}(r) = &\;\dfrac{q\, r}{2 r_0}\left( 3 - \dfrac{r^2}{r_0^2} \right) \nonumber \\ &\;- \left[\dfrac{3q}{r_0}\left(1- \dfrac{r^2}{r_0^2}\right)\right] \frac{r^2+a^2}{a} \arctan \frac{a}{r},\label{eq:tspherq}
\end{align}
where we used relations \eqref{eq:electen} and \eqref{eq:Q-(r)}.

Naturally, since the exterior current density $J_+^{\mu}$ vanishes, the total electric charge inside a spheroid of $r = \text{constant}\geq r_0$ may be obtained directly from \eqref{eq:tspherq} by taking the limit $r\to r_0$, which gives
\begin{align}
    \mathfrak{q}_{+}(r) = q. \label{eq:tspherq+}
\end{align}

Now we have all the ingredients to present the complete solution for the two spacetime regions, what is done in the next section.

\subsection{The complete rotating  geometry and its main properties}
\label{sec:exsol1}

\subsubsection{The interior geometry}
\label{sec:rintsol}

In the Boyer-Lindquist coordinates, the interior solution is characterized by the G\"urses-G\"ursey metric~\eqref{eq:ggmetric} with the mass function $M_-(r)$ given by Eq.~\eqref{eq:m-(r)}.

In addition, the electromagnetic gauge potential together and the electromagnetic strength tensor are given by Eqs.~\eqref{eq:rotpot} and~\eqref{eq:electen}, respectively, with $Q_-(r)$ given in Eq.~\eqref{eq:Q-(r)}. The resulting gauge potential reads
\begin{align}
\mathcal{A}_{-\mu} = \dfrac{q}{2 r_0\Sigma}\left( 3 - \dfrac{r^2}{r_0^2} \right)\left(\delta_{\mu}^{\ t} - a \sin^2 \theta \delta_{\mu}^{\ \varphi}\right), \label{eq:rotpot-} 
\end{align}
and the Faraday-Maxwell strength tensor reads
\begin{align}
    & F_{- rt} =  \dfrac{q\, r}{2 r_0}\left( 3 - \dfrac{r^2}{r_0^2} \right)\frac{r^2 - a^2 \cos^2 \theta}{\Sigma^2} -\dfrac{3q\, r}{2r_0\Sigma}\left(1- \frac{r^2}{r_0^2} \right), \nonumber \\
    & F_{- \theta t} = -  \dfrac{q\, r}{2 r_0}\left( 3 - \dfrac{r^2}{r_0^2} \right) \frac{ r\, a^2 \sin 2 \theta}{\Sigma^2}, \nonumber \\
    & F_{- r\varphi} = - a \sin^2 \theta F_{- rt}, \label{eq:electen-} \\ 
    & F_{-\theta \varphi} = -\frac{\left(r^2+a^2\right)}{a} F_{-\theta t}.  \nonumber
\end{align}
The current density that generates the interior electromagnetic field \eqref{eq:electen-} is given by
\begin{align}
      4\pi J_-^t = & \frac{3q}{r_0}\frac{r^2 +a^2}{\Sigma^3}\left[r^2 -\left( 1- \frac{2r^2}{r_0^2} \right)a^2\cos^2\theta \right] \nonumber \\ &- \frac{3q}{r_0}\frac{r^2}{\Sigma^2}\left(1 -\frac{r^2}{r_0^2}\right),\\
     4 \pi J_-^{\varphi} = & \frac{3q}{r_0}\frac{a}{\Sigma^3}\left[r^2 -\left( 1- \frac{2r^2}{r_0^2} \right)a^2\cos^2\theta \right],
\end{align}
with the other components being identically zero. The current density in the exterior region vanishes, $J_+^\mu =0$.
It is immediately seen that, for vanishing rotation $a\to 0$, the only nonzero component of the current density is $J_-^t = {3q}/(4\pi r_0^3) $, exactly the result for the static case [see Eq.~\eqref{eq:stJ-}]. Moreover, the components of the current density are finite everywhere in the interior region, with the exception of the ring $(r=0,\, \theta=\pi/2)$, where it diverges. Inside the ring, at the disc $(r=0,\, 0\leq \theta<\pi/2)$,  the components of the current density are finite and given by $J^t = -3q\big/\!\left(r_0 a^2 \cos^4 \theta\right)$ and $J^{\varphi} = -3q\big/\!\left(r_0 a^3 \cos^4 \theta\right)$, i.e., $J^{t}=aJ^{\varphi}$. On the other hand, the electromagnetic field tensor given in \eqref{eq:electen-} is finite everywhere inside the fluid and vanishes on the disk and on the ring.

From the decomposition of the current density, as measured by a comoving observer with the fluid, presented in Appendix \ref{sec:appendixA}, cf. Eqs.\eqref{eq:jmu-}--\eqref{eq:conduction-}, it follows that the convective charge density and the current density are given by
\begin{align}
   & 4\pi  \rho_{e} = \sqrt{\frac{\pm \Delta_-}{\Sigma}}\frac{3q}{r_0\Sigma^2}\!\left[\frac{r^4}{r_0^2} -\!\left( 1- \frac{2r^2}{r_0^2} \right)\!a^2\cos^2\theta \right], \label{eq:chargedens-} \\
   & 4 \pi j^{\mu} = \frac{3q}{r_0} \frac{r^2}{\Sigma^2}\left(1 - \frac{r^2}{r_0^2} \right) \frac{a \sin \theta}{\sqrt{\Sigma}}e_3^{\ \mu}, \label{eq:conducdens-}
\end{align}
respectively, with $e_3^{\ \mu}$ being the fourth component of the Carter tetrad. The convective charge density~\eqref{eq:chargedens-} and the current density~\eqref{eq:conducdens-} are finite everywhere with the exception of the ring $(r=0,\, \theta=\pi/2)$, where it diverges. Inside the ring, at the disc $(r=0,\, 0\leq \theta<\pi/2)$, the convective charge density is finite and given by $4\pi \rho_{e} = -3q/\left(r_0 a^2 \cos^3 \theta\right)$ while the current density vanishes. At the boundary, one gets $4\pi \rho_e= 3q \sqrt{\pm \Delta_-(r_0)/ \Sigma_0}\,/\left(r_0 \Sigma_0\right)$ and $j^{\mu} = 0$ with $\Sigma_0 = r_0^2 + a^2 \cos^2 \theta$.

\subsubsection{The exterior geometry}

In the Boyer-Lindquist coordinates, the exterior solution is characterized by the G\"urses-G\"ursey metric~\eqref{eq:ggmetric} with the mass function $M_+(r) = m - q^2/2r$, cf. Eq.~\eqref{eq:m+(r)}.

The electromagnetic gauge potential and the electromagnetic strength tensor are given by Eqs.~\eqref{eq:rotpot} and~\eqref{eq:electen}, respectively, with $Q_+(r) = q$. Therefore, the exterior solution is exactly the well-known Kerr-Newman spacetime.

\subsubsection{Matching conditions}

\label{sec:matchr}
Similarly to the static case, the smooth junction between the Kerr-Newman exterior metric and the G\"urses-G\"ursey interior metric found in Sec.~\ref{sec:rintsol} is performed by employing the matching conditions of Darmois-Israel~\cite{Israel66}.

Following the work of Drake and Turolla~\cite{Drake97}, we first choose the boundary surface $\mathcal{B}_r$ that separates the interior region from the exterior region. The simplest choice is $\mathcal{B}_r\!:\! r = r_0 = \text{constant}$. This implies that the interior solution is defined by the values $r < r_0$ while the exterior region is defined in the interval $r_0 < r < \infty$. The possible extension of the interior metric to negative values of $r$ is not investigated in the present work.  Following the same steps of Sec.~\ref{sec:match}, 
we define $\xi^a = (\tau, \, \theta,\, \varphi)$ to be the intrinsic coordinates of the surface $\mathcal{B}_r$ and $x^{\mu}_{\pm} = (t,\, r, \, \theta, \, \varphi)$ to be the coordinates of both the interior region $\mathcal{M}_-$ and the exterior region $\mathcal{M_+}$, respectively. As in the static case dealt with in Sec.~\ref{sec:match}, due to the symmetry of the geometry, and since we are considering smooth boundary conditions, the coordinates of the two spacetime regions may be identified, i.e., $t_-=t_+=t$, $r_-=r_+=r$, $\theta_-=\theta_+=\theta$, and $\varphi_-=\varphi_+=\varphi$. Hence, from now on we drop the $\pm$ indexes.  

Once again, we notice that the assumption of a smooth transition across the boundary implies that the first and second fundamental forms are continuous across $\mathcal{B}_r$, i.e., $[h_{ab}] = 0$ and $[K_{ab}] = 0$. Therefore, by applying the smooth matching conditions, as found in~\cite{Masa22}, we get
\begin{align}
    & M_-(r_0) = M_+(r_0), \label{eq:match3} \\
    & M_-^\prime(r)\big|_{r = r_0} = M_+^\prime(r)\big |_{r = r_0}.
 \label{eq:match4}
\end{align} 
Interestingly, Eqs.~\eqref{eq:match3} and \eqref{eq:match4} imply the same matching conditions found in the static case, i.e., the result is the same as given in Eqs.~\eqref{eq:match1} and \eqref{eq:match2}.
Moreover, the matching conditions~\eqref{eq:match3}, \eqref{eq:match4} together with Eq.~\eqref{eq:qr0},  also imply that the boundary surface $\mathcal{B}_r$ is defined by $r=r_0 = \sqrt{2/3}b$, a valid relation in the static case as well, cf. Eq.~\eqref{eq:br0R}.

In addition, the electromagnetic fields are smoothly matched together at the surface $\mathcal{B}_r$, i.e, $[A_a] = 0$, $[F_{ab}] = 0$ and $[F_{an}] = 0$ (see Sec.~\ref{sec:match}). From Eqs.~\eqref{eq:rotpot} and \eqref{eq:electen}, these smooth matching conditions lead to 
\begin{align}
    & Q_-(r_0) = Q_+(r_0)=q, \label{eq:match5} \\
    & Q_-^\prime(r)\Big|_{r = r_0} = Q_+^\prime(r)\Big |_{r = r_0} =0,
 \label{eq:match6}
\end{align}
also the same result as in the static case.

Finally, it is easy to see from Eq.~\eqref{eq:tspherq} that the total charge inside the spheroid defined by $r = r_0$ is $\mathfrak{q}_-(r_0) = q= \mathfrak{q}_+(r_0)$. In fact, this is the net charge inside any spheroid defined $r = {\rm constant}\geq r_0$. Therefore, we observe that the total electric charge of the static solution is preserved, even though the current density of the rotating solution is quite different from the static solution.

\subsubsection{Curvature regularity}
\label{sec:curvreg}
For the Kerr spacetime, where the mass function $M_{\pm}(r) = m$ is a constant, and for the Kerr-Newmann spacetime, where the mass function is given by $M_{\pm}(r)= m - q^2/2r$, it is well-known that the region defined by $S^1\!\!:\!(r = 0,\, \theta = \pi/2)$ corresponds to a ring curvature singularity. On the other hand, as it was shown by Torres~\cite{Torres} and Maeda~\cite{Maeda}, if the mass function $M_-(r)$ can be expanded around $r = 0$ as $M_-(r) \approx c_0 r^{3 + \epsilon}$ with $c_0 \neq 0$ and $\epsilon \ge 0$, then the ring  $S^1$ corresponds to a conical singularity (when extended to the region $r<0$), and not to a scalar polynomial curvature singularity. On the other hand, if the extension through the region $r < 0$ is not considered, the conical singularity is also absent, as shown recently by Torres~\cite{Ramon23}.

In the present case, for the interior region, where the ring is located, the mass function is given by Eq.~\eqref{eq:m-(r)}, which implies that, around $r = 0$, we have $M_-(r) \approx r^3/2R^2$ and, therefore, the ring $S^1$ does not correspond to a curvature singularity. Moreover, since we do not perform the extension through the region $r<0$ in this work, the ring $S^1$ does not correspond to a conical singularity as well. Nevertheless, it is interesting to mention that, even in the zero mass limit of the Kerr metric, $S^1$ is a ring-like conical singularity when the coordinate $r$ is extended to negative values, and therefore the zero mass limit of the Kerr metric corresponds to a wormhole spacetime \cite{Gibbons}.

Moreover, as already noticed in several works~\cite{Spallucci, Modesto, Torres, Maeda, Masa22}, even in the cases when the curvature scalars  assume finite values at that ring, they s are not well-defined there for $\epsilon = 0$. This also happens in the present case. To see this, let us consider just the Ricci scalar for the G\"urses-G\"ursey metric \eqref{eq:ggmetric}, i.e., 
\begin{align}
    \mathcal R = \frac{2}{\Sigma}\left[2M_-'(r) + r\,M_-''(r)\right],
\end{align}
which for the present interior solution, with $m_-(r)$ given by \eqref{eq:m-(r)}, reduces to
\begin{align}
    \mathcal R = \frac{12 r^2}{R^2\Sigma}\left(1 - \frac{r^2}{r^2_0}\right). 
\end{align}
Then one can see that, for $\theta = \pi/2$, the limit $r \to 0$ gives $\mathcal R = 12/R^2$, whereas, for $\theta \neq \pi/2$, the limit $r \to 0$ gives $\mathcal{R} = 0$. Hence, there is a finite jump in the Ricci scalar, what also happens with the other independent curvatures scalars (see e.e. \cite{Torres}).  This property presented by rotating solutions of G\"urses-G\"ursey type with nonsingular scalar curvature has been interpreted in \cite{Spallucci} as a planar string replacing the ring singularity of the Kerr solution.

\subsubsection{The energy-momentum tensor}
\label{sec:emt}

The general expression for the total EMT of the rotating solution and its decomposition in terms of the Carter's tetrad is presented in Appendix~\ref{sec:appendixEMT}. 
Here we use those expressions to examine the energy-momentum tensor (EMT) resulting from the complete rotating solution presented above.  

By using Carter's orthonormal tetrad defined in Eq.~\eqref{eq:rtetrad}, the total EMT assumes the diagonal form $T_{\mu\nu}e^\mu_a e^\nu_b = {\rm diag}\left(\varrho,\, \mathfrak p_1,\, \mathfrak p_2,\,  \mathfrak p_3\right)$,
with the eigenvalues $\varrho$ and $\mathfrak p_1$ corresponding to the total energy density and to the effective radial pressure, respectively. The eigenvalues $\mathfrak p_2$ and $\mathfrak p_3$ are the effective tangential pressures. 
From Eqs.~\eqref{eq:rpressa} and \eqref{eq:tpressa}, together with the mass functions \eqref{eq:m-(r)} and \eqref{eq:m+(r)}, we obtain the explicit expressions for the eigenvalues. Namely,
\begin{align}
     &8\pi \varrho_- = -8\pi \mathfrak p_{1-} =\frac{3r^2}{R^2}\frac{r^2}{\Sigma^2}\left(1 - \frac23\frac{r^2}{r_0^2}\right) , \label{eq:totalrpress-}\\
     & 8\pi \mathfrak p_{2-} = 8\pi \mathfrak p_{3-} = 8\pi \varrho_-  - \frac{6 r^2}{R^2 \Sigma}\left(1 - \frac{r^2}{r_0^2}\right) \label{eq:totaltpress-}.
\end{align} 
Notice that the energy density and the pressures present a similar behavior at $r\to 0$ as the curvature scalars analyzed in the previous section.

The eigenvalues defined in the last two equations carry also the electromagnetic contribution $E_{\mu\nu}$. 
 In the Carter's orthonormal frame, the eigenvalues of the electromagnetic EMT are obtained by the usual relations $E_{\mu\nu}e_a^\mu e _b^\nu$, as defined in Appendix \ref{sec:appendixEMT}. Using Eqs.~\eqref{eq:emtEMa},~\eqref{eq:electen} and~\eqref{eq:Q-(r)}, we get 
\begin{align}
  \!  & 8\pi\varrho_{em}
    = \frac{q^2 r^6}{\Sigma^2 r_0^6} + \frac{3q^2 r^2 a^2 \cos^2\theta}{\Sigma^3 r_0^2}\left[3 - 4\frac{r^2}{r_0^2} + \frac{r^4}{r_0^4} \right], \label{eq:emtEM00}\\ 
  \!  & \mathfrak p_{em 1}
    =-\mathfrak p_{em 2}
    = -\mathfrak p_{em 3}
    = -\varrho_{em},  \label{eq:emtEM22}
\end{align}
which we can see that the energy density and pressures of the electromagnetic field vanishes independent of the path taken to $r = 0$.

As in the static case, the total EMT can be split in terms of the matter EMT and of the electromagnetic EMT, i.e., $T_{\mu\nu} = M_{\mu\nu} + E_{\mu\nu}$, and then we extract the matter contribution to the energy density and pressures by subtracting $E_{\mu\nu}$ from $T_{\mu\nu}$.
Therefore, in the Carter's frame, the eigenvalues of the matter EMT are given by
\begin{align}
  8\pi \varrho_{m} = \; & - 8\pi \mathfrak p_{m 1} = \frac{15mr^4}{4 r^3_0\Sigma^2}\left(1 - \frac{r^2}{r_0^2}\right)\nonumber \\ & - \frac{15mr^2a^2\cos^2\theta}{4 r_0 \Sigma^3}\left(3 - 4\frac{r^2}{r_0^2} + \frac{r^4}{r_0^4} \right),   \\
8\pi  \mathfrak p_{m 2} & =  8\pi\mathfrak p_{m 3} = 8\pi \varrho_{m} -  \frac{15mr^2}{2 r_0^3 \Sigma}\left(1 \;- \frac{r^2}{r_0^2}\right),
\end{align}
which represents an anisotropic fluid with energy density and pressures vanishing at boundary $r = r_0$. 
Notice that the matter EMT vanishes at $r=r_0$. It is also noteworthy that the energy density and pressures of matter present the same behavior at the ring $r\to 0$ as the curvature scalars analyzed in the previous section.

Finally, by using Eq.~\eqref{eq:tpressa}, together with the mass function \eqref{eq:m+(r)}, we obtain the explicit expressions for the eigenvalues valid in the exterior region $\mathcal {M}_+$. Namely,
\begin{align}
       &8\pi \varrho_+ = -8\pi \mathfrak p_{1+} = \frac{3}{R^2}\frac{r_0^4}{\Sigma^2}= \frac{15mr^4}{4r_0^3\Sigma^2}=\frac{q^2}{\Sigma^2}, \label{eq:totalrpress+}\\
   &8\pi \mathfrak p_{2+}= 8\pi \mathfrak p_{3+} = 8\pi \varrho_+ \label{eq:totaltpress+},
\end{align} 
that is due to the presence of the electromagnetic vacuum field. 
Then, by comparing Eqs.~\eqref{eq:totalrpress-} and \eqref{eq:totaltpress-} with Eqs.~\eqref{eq:totalrpress+} and \eqref{eq:totaltpress+}, it is seen that the EMT is continuous across the boundary. 
In fact, since the matter EMT vanishes at the boundary, the continuity of the total EMT is guaranteed by the electromagnetic EMT.

\subsubsection{Energy conditions}
\label{sec:energycond}

 It was shown by Torres \cite{Torres} that, if the mass function of the G\"urses-G\"ursey metric $M_-(r)$ can be expanded in a Taylor polynomial series around $r = 0$, then the WEC should be violated around $r = 0$. In addition, Maeda~\cite{Maeda} extended the result of Torres by showing that if the mass function $M_-(r)$ of the G\"urses-G\"ursey metric can be expanded around the locus $r = 0$, with $\theta \neq \pi/2$, as $M_-(r) \approx c_0 r^{3 + \epsilon}$ with $\epsilon \ge 0$, then all energy conditions are violated close to $r = 0$ for $c_0 > 0$ and, while the NEC and the SEC are respected for $c_0 < 0$. Therefore, since close to $r=0$ the mass function \eqref{eq:m-(r)} is such that $M_-(r) \approx r^3/2R^2$, which gives $c_0 = R^{-2}/2 > 0$, it is straightforward to see that all the energy conditions are violated around the region $r = 0$, $\theta \neq \pi/2$.  Interestingly, for $\theta = \pi/2$, the energy conditions are identical to those presented by the original static spacetime that seeds the rotating geometry.

\subsubsection{On the horizons, ergosurfaces, and other properties}
\label{sec:horiprop}

We start by investigating the existence of horizons in the G\"urses-G\"ursey metric~\eqref{eq:ggmetric}. As well known, is such a case, the horizons are located at the loci where the function $\Delta(r)$ given by Eq.~\eqref{eq:deltfun} vanishes. The explicit forms of the equations for such loci are obtained by replacing Eqs.~\eqref{eq:qr0R} and~\eqref{eq:mr0Rq} into Eq.~\eqref{eq:deltfun}. 
Then, it follows that the horizons radii are the real and positive roots of the polynomial equations 
\begin{align}
&  \Delta_-(r) = r^2 + a^2 - \frac{5m r}{4}\left(\frac{r}{r_0} \right)^{\!3} \left(1 - \frac{2r^2} {5r^2_0} \right) = 0, \label{eq:horieq1}  \\
&  \Delta_+(r) = r^2 + a^2 - 2mr + \frac{5 m r_0}{4} = 0. \label{eq:horieq2}
\end{align}
These two equations must be supplemented by two additional conditions. Namely, Eq.~\eqref{eq:horieq1} defines the horizons if its roots are in the interval $r \le r_0$, while the horizons are defined by the roots of Eq.~\eqref{eq:horieq2} when such roots are in the interval $r\geq r_0$. The roots of both equations may be expressed in the form of radicals, but we do not write them here to avoid cumbersome expressions.

\begin{figure}[tbh]
    \centering
        \centering
        \includegraphics[width=7.95cm]{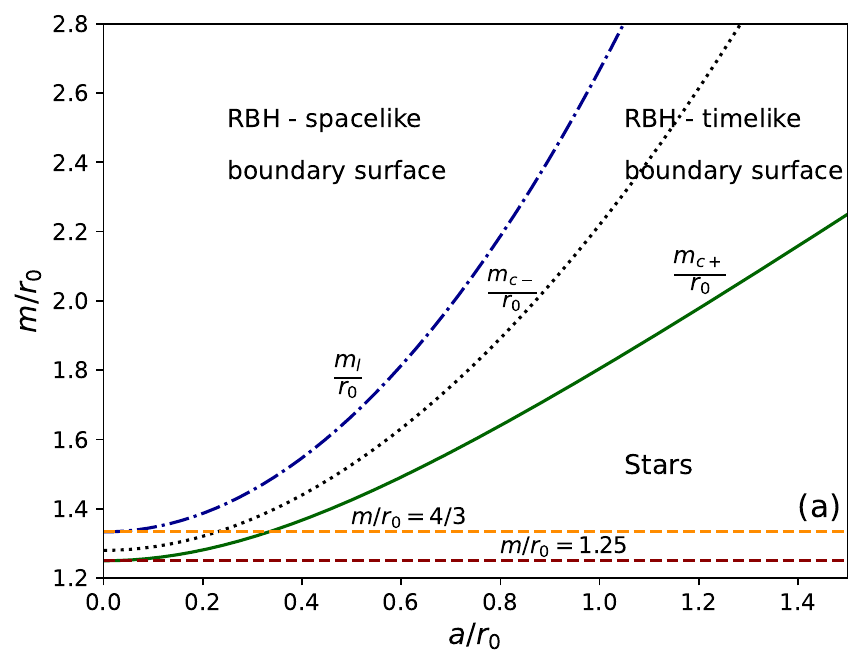}
        \centering
        \includegraphics[width=8.2cm]{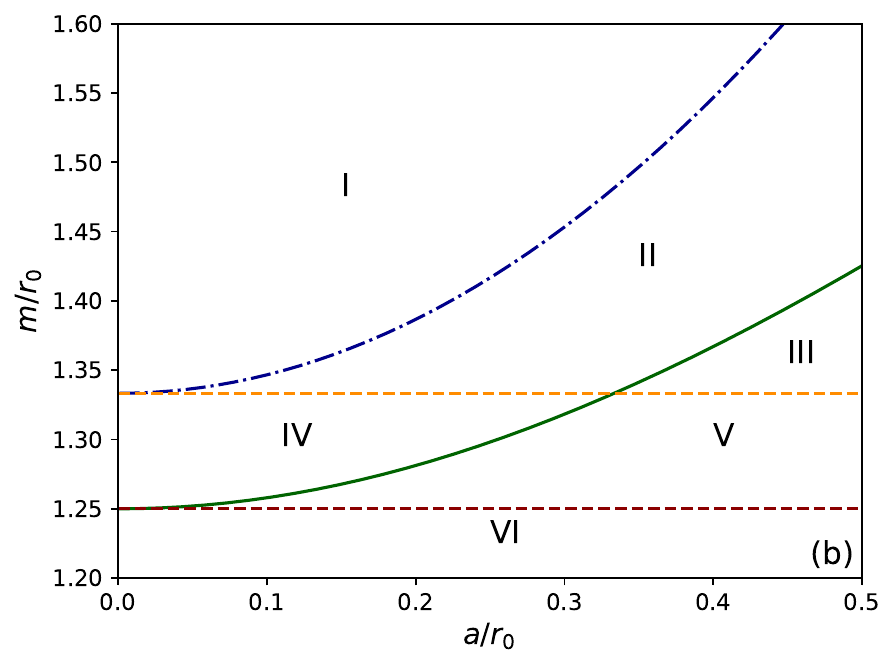}
    \caption{Panel (a), top: A plot of the limiting and extremal masses as a function of $a/r_0$ in the parameter space spanned by $m/r_0$ and $a/r_0$. These mass functions are important to determine the different kinds of objects and the properties of boundary surfaces. The respective minimum constant values $m_l/r_0=4/3$ and $m_{c+}/r_0=m_{c-}/r_0=32/25$ are also drawn, as indicated by the respective labels. Such minimum values as for determining the properties of the ergosurfaces. The properties of the different kind of objects indicated in the figure are described in the main text (see Sec.~\ref{sec:objects}). \\
    Panel (b), bottom: The region $0 < a/r_0 < 0.5$ of the top panel zoomed in, aiming to display the six different regions of the space of parameter containing different kinds of objects also regarding the ergosurfaces. 
    Regions I, II, and IV contains regular black holes, while regions III, V, and VI contains star-like objects, i.e., configurations with no horizons. 
    }
    \label{fig:mr0}
\end{figure}

A simple analysis shows that the polynomial $\Delta_-(r)$ may have at most two real positive roots, which we indicate by $r_{1-}$ and $r_{2-}$, with $r_{1-}\leq r_{2-}$. Similarly, the polynomial $\Delta_+(r)$ may have at most two real positive roots, which we indicate by $r_{1+}$ and $r_{2+}$, with $r_{1+}\leq r_{2+}$.
Hence, by following the same reasoning of Ref.~\cite{Dymn15}, it is seen that each one of Eqs.~\eqref{eq:horieq1} and~\eqref{eq:horieq2} can present two, one, or none real positive roots, depending on the relative values of the normalized free parameters $m/r_0$ and $a/r_0$.
The critical case that separates a spacetime with no horizons from a spacetime with two horizons is the extremal case, in which the two real positive roots of the horizon function $\Delta_-(r)$ or $\Delta_+(r)$ become identical. i.e.,  $r_{1-} = r_{2-}$ or $r_{1+} = r_{2+}$. 
This situation is equivalent to a vanishing discriminant of the corresponding polynomial function $\Delta_-(r)$ or $\Delta_+(r)$, what in turn yields the extremal masses $m_{c\pm}$, respectively. The expressions for these extremal masses in terms of $a$ and $r_0$ are
\begin{align}
\frac{m_{c-}}{r_0} &\; =  \frac{1}{250}\Bigg[180 + 108 \frac{a^2}{r_0^2} - 25\frac{r_0^2}{a^2}\label{eq:discri1} \\ & + \sqrt{\left(180 + 108 \frac{a^2}{r_0^2} - 25\frac{r_0^2}{a^2}\right)^2 + 500\times32\frac{r_0^2}{a^2}} \Bigg], \nonumber
\end{align}
for $\Delta_-(r)$ , and
\begin{align}
    \frac{m_{c+}}{r_0}  = \frac{5}{8}\left(1 + \sqrt{1 + \frac{64}{25}\frac{a^2}{r_0^2}} \right) \label{eq:discri2},
\end{align}
for $\Delta_+(r)$. 

Let us first notice that, the three free parameters of the model appearing in these relations, namely, $m$, $a$, and $r_0$, combine in such a way that only their normalized relative values really matter. Therefore, without loss of generality, we may normalize the mass and the rotation parameter by $r_0$, and use $m/r_0$ and $a/r_0$ as the actual free parameters of the model.  

Relations \eqref{eq:discri1} and \eqref{eq:discri2} mean that, given a fixed value of the normalized rotation parameter $a/r_0$, configurations bearing masses smaller than $m_{c\pm} $ present no real roots, configurations bearing masses larger than $m_{c\pm}$ present two real roots, while configurations bearing masses $m$ such that $m=m_{c\pm}$ present one real double root. 

Conversely, we may invert the last relations and express the rotation parameter in terms of the mass, thus obtaining extremal values for the interior and exterior solutions, $a_{c-}$ and $a_{c+}$, respectively. Since the expressions for $a_{c\pm}$ plays similar role to the extremal masses we do not write them here.  In fact, the interpretation of such extremal values of $a_{c\pm}$ is also similar to the extremal masses.
Namely, for a given value of the normalized mass $m/r_0$, configurations with rotation parameters smaller than $a_{c\pm}$ present two real roots,
configurations bearing rotation parameters larger than $a_{c\pm}$ present no real roots, corresponding to overextremal solutions, while configurations bearing rotation parameters such that $a=a_{c\pm}$ present one real double root, corresponding to extremal solutions. Therefore, the analysis of the roots of the coefficients $\Delta_\pm(r)$ and the possible presence of horizons may be performed just by using relations \eqref{eq:discri1} and \eqref{eq:discri2}, as we do in the remaining of the present section.

The situation here is similar to what happens regarding the existence of horizons in the static solution presented in the previous section. For instance, the larger root of $\Delta_{-}(r)$, $r_{2-}$, is larger than the radius of the boundary surface for all values of the parameters $m/r_0$ and $a/r_0$ that yield real roots. As a consequence, the spacetime region $\mathcal{M}_-$ possesses at most one horizon that corresponds to configurations for which the root $r_{1-}$ belongs to the spacetime region $\mathcal M_-$.

In order to the interior spacetime region to actually present horizons, at least one of the zeros of $\Delta_-(r)$ must be located in the region $r/r_0 \le 1$, while for the exterior spacetime to actually have horizons, at least one of the zeros of $\Delta_+(r)$ be located in the region $r/r_0 \ge 1$. In the limiting case, one of the roots of $\Delta_-(r)$ shall coincide with the boundary surface, i.e., $\Delta_-(r = r_0)= 0$. Similarly, in the limiting case, one of the roots of  $\Delta_+(r)$ shall coincide with $r_0$, i.e., $\Delta_+(r = r_0)=0$. These two conditions yield the same relation $r_0^2 + a^2 -3m\, r_0/4=0$, which furnishes the limiting relation
\begin{align}
   \frac{m_l}{r_0} = \frac{4}{3}\left(1 + \frac{a^2}{r_0^2} \right),\ \ {\rm or}\ \ \frac{a_l^2}{r_0^2} = \frac34\frac{m}{r_0} -1.\label{eq:ml}
\end{align}
Equation \eqref{eq:ml} means that,  for masses smaller than $m_l$, at least one of the roots of $\Delta_+(r)$ is larger $r_0$, defining a horizon in the vacuum region of spacetime. On the other hand, for masses larger than $m_l$, at least one of the roots of $\Delta_-(r)$ is smaller than $r_0$, indicating a horizon in the matter region of the spacetime. Moreover, since the larger root of $\Delta_-(r)$ is larger than $r_0$, the spacetimes corresponding to this limiting mass present the Cauchy horizon which coincides with the boundary surface at $r=r_0$.

Panels (a) and (b) of Fig.~\ref{fig:mr0} show the behavior of the normalized masses $m_{c-}/r_0$, $m_{c+}/r_0$, and $m_l/r_0$ as a function of the normalized rotation parameter $a/r_0$. The first fact to notice is that $m_l$ is larger than the other extremal masses, i.e., $m_{c+}/r_0 < m_{c-}/r_0 < m_l/r_0$, for all values of $a/r_0$ that yield real values for $m_{c\pm}$. Other aspect to be mentioned is that the three masses grow monotonically with the rotation parameter. The minimum values of such masses are $m_{c+}/r_0=5/4$, $ m_{c-}/r_0=32/25 $, and $ m_l/r_0= 4/3$, that are found at $a/r_0=0$.

\begin{figure}[t!]
    \centering
        \centering
        \includegraphics[width=8.5cm]{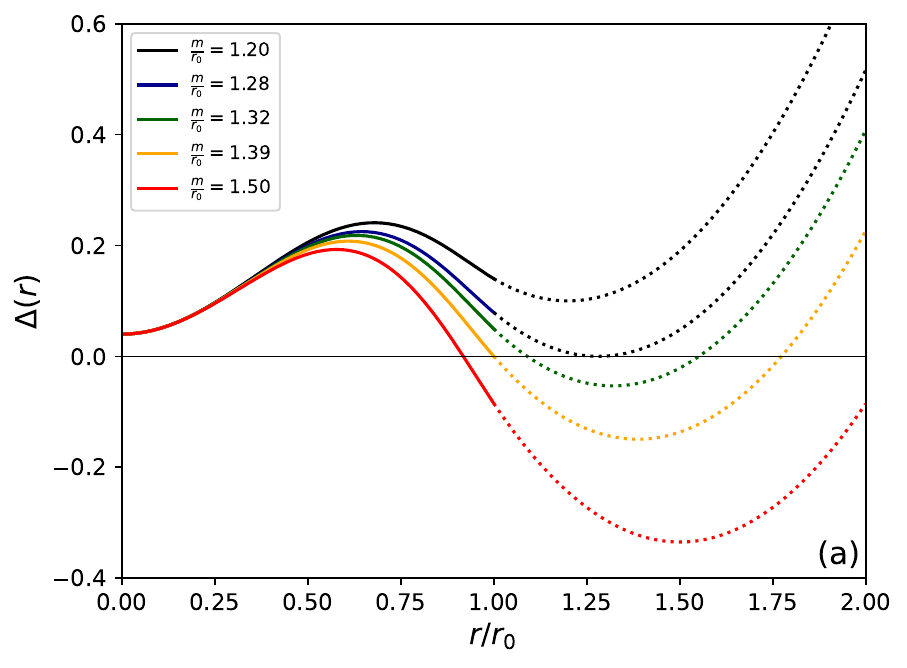}
        \includegraphics[width=8.5cm]{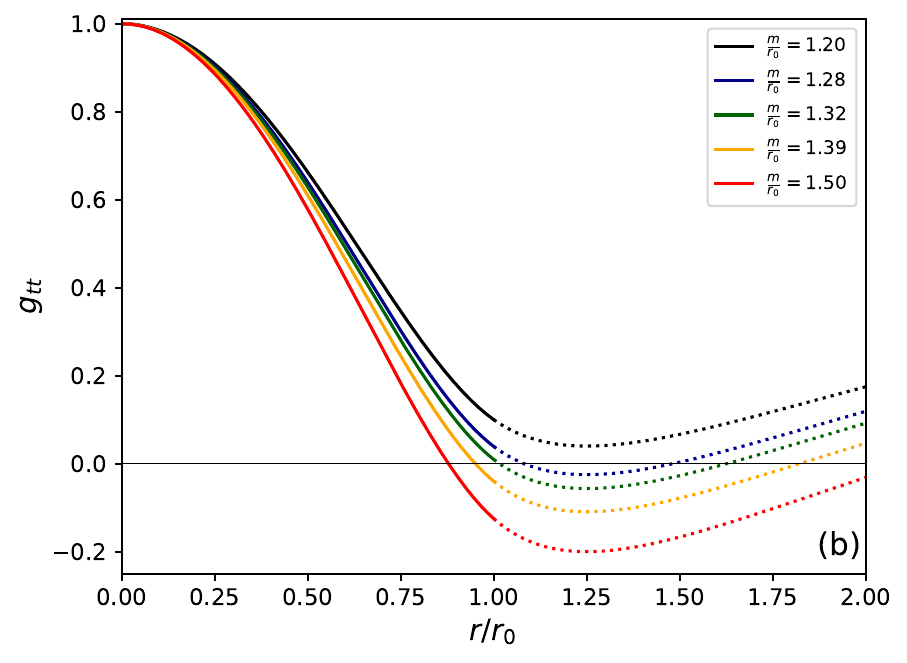}
        
    \caption{Panel (a), top: The horizon function as a function of ${r}/{r_0}$, for different values of the total mass $m/r_0$ (as indicated in the legend) with fixed $a/r_0 = 0.2$. In this case, the extremal mass $m_{c+}$  is such that $m_{c+}/r_0 \simeq 1.28$, while the limiting mass $m_l$ is such that $m_l/r_0 \simeq 1.39$. The solid lines represent the interior solution  $\Delta_-(r)$, while the dotted lines represent the corresponding exterior solution $\Delta_+(r)$. The zero of $\Delta(r)$ indicates the presence and the location of the horizons.\\
    Panel (b), bottom: The metric function $g_{tt}$ as a function of ${r}/{r_0}$ on the transverse plane $\theta = \pi/2$, for different values of the total mass $m/r_0$ (as indicated in the legend) with fixed $a/r_0 = 0.2$. In this case, the extremal mass $m_{c+}$ is such that $m_{c+}/r_0 \simeq 1.28$ while the limiting mass $m_l$ is such that $m_l/r_0 \simeq 1.39$. The solid lines represent the interior solution, while the dotted lines represent the corresponding exterior solution. The zeros of $g_{tt}$ indicate the presence of ergosurfaces.}
    \label{fig:rthorizons}
\end{figure}

Panel (a) of Fig.~\ref{fig:rthorizons} shows the behavior of the horizon function $\Delta(r)$ as a function of the normalized radial coordinate ${r}/{r_0}$, for four different values of the total mass $m/r_0$, and with a fixed normalized rotation parameter, $a/r_0 = 0.2$.
Another interesting property of a spacetime generated by a rotating source is the existence of the stationary limit surfaces, the so-called ergosurfaces. Such surfaces form the boundary of the ergoregions, and are determined by the causal character of the stationary Killing vector $\xi^{\mu} = \delta^{\mu}_{\ t}$, being located in the spacetime regions where $\xi^{\mu}$ becomes lightlike. This condition implies that $g_{tt} = 0$ which, in the present case, implies the following equations
\begin{align}
& r^2 + a^2\cos^2 \theta - \frac{5m r}{4}\left(\frac{r}{r_0} \right)^3 \left(1 - \frac{2r^2} {5r^2_0} \right) = 0, \ r \le r_0,  \nonumber  \\
& r^2 + a^2\cos^2 \theta - 2mr + \frac{5 m r_0}{4} = 0, \ r\ge r_0. \label{eq:ergoeq2}
\end{align}
Similarly to the equations that determines the presence of horizons, these equations can have two real positive roots that define two ergosurfaces.

Panel (b) of Fig.~\ref{fig:rthorizons} shows
the metric function $g_{tt}$ on the transverse plane $\theta = \pi/2$, for the same values of parameters as in panel (a) of Fig.~\ref{fig:rthorizons}.

By inspecting panels (a) and (b) of Fig.~\eqref{fig:rthorizons}, we may draw some interesting assertions about horizons and ergosurfaces of configurations with rotation parameter such that $a/r_0=0.2$. 

Configurations with mass $m/r_0 = 1.20$ present no horizons nor ergosurfaces, and the corresponding spacetimes represent charged rotating regular star-like objects with a timelike boundary. 

Configurations with mass $m/r_0 = 1.28$ present a degenerate horizon and two ergosurfaces all located in the exterior region, outside matter, representing extremal charged rotating regular black holes with a core of charged matter that has a timelike boundary. 

Configurations with mass $m/r_0 = 1.32$ present two horizons and two ergosurfaces all located in the exterior region, outside matter, representing charged rotating regular black holes with a core of charged matter that has a timelike boundary. 

Configurations with the limiting mass  mass $m/r_0 = m_{c+}/r_0= 1.39$, presents two horizons with the Cauchy horizon located at the boundary of the matter distribution, the event horizon is outside matter, while the ergosurfaces extend from inside to outside matter. This kind of configurations represents charged rotating regular black holes with a central core of matter that has a lightlike boundary. 

Finally, configurations with mass $m/r_0 = 1.50$ present two horizons with the Cauchy horizon being inside matter, with the event horizon being outside matter, and with the ergosurfaces extending from inside to outside matter. The boundary surface is located between the two horizons and then this kind of configurations represents charged rotating regular black holes with a central core of matter that has a spacelike  boundary.

The aspects just evidenced in regard to the particular cases depicted in panels (a) and (b) of Fig.~\ref{fig:rthorizons} hold in general. In fact,  we may draw the following conclusions.  \vskip .1cm
\noindent
{\it i}) The extremal mass $m_{c-}$ is not relevant for the present analysis, since it furnishes a double root of the interior function $\Delta_-(r)$ which is larger than the boundary surface radius, i.e., $r_{k-}> r_0$, and then it does not represent any actual configuration. 
\vskip .1cm
\noindent
{\it ii}) The extremal mass $m_{c+}$ is the one that will separate configurations with no horizons from configurations with two horizons.\vskip .1cm
\noindent
{\it iii}) The limiting mass $m_l$ gives the condition for the boundary surface $r_0$ to coincide with one of the horizons, and then it also tell us about the causal nature of the matching condition.\vskip .1cm
\noindent
{\it iv}) Configurations with masses in the interval $0< m/r_0 < m_{c+}/r_0$ present no horizons, corresponding to over-extremal configurations. \vskip .1cm
\noindent
{\it v}) 
Configurations with masses in the interval $m_{c+}/r_0< m/r_0 < m_l/r_0$ present two horizons, a Cauchy horizon in the exterior region and an event horizon in the exterior region.  \vskip .1cm
\noindent
{\it vi}) 
Configurations with masses such that $m/r_0 = m_l/r_0$ present two horizons, a Cauchy horizon at the boundary surface, and an event horizon in the exterior region.  \vskip .1cm
\noindent
{\it vii}) Configurations with very large masses, i.e., with masses satisfying the inequality $m/r_0 > m_l/r_0$, present two horizons, a Cauchy horizon located in the interior region and tan event horizon located in the exterior.  

A more detailed description of the different classes of objects represented by the complete rotating solution are presented next.

\subsection{The different kinds of objects in the complete rotating geometry}
\label{sec:objects}

\subsubsection{Nonextremal rotating regular black holes with a spacelike  boundary - Region I}

Configurations with very large masses, characterized by the inequality $m/r_0 > m_l/r_0$, present two horizons, a Cauchy horizon located inside matter and given by the smallest positive root of $\Delta_-(r)$, $r_{1-}$, and an event horizon located outside matter and given by the largest root of $\Delta_+(r)$,  $r_{2+}$. Such configurations present also two ergosurfaces, with the inner ergosurface being located completely inside matter, while the outer ergosurface is located completely outside matter. The boundary of the matter distribution is located between the two horizons, being a spacelike surface. Hence, the corresponding spacetimes are nonextremal rotating regular black holes whose central core of matter has a spacelike boundary. This situation corresponds to configurations whose parameters belong to the region above the dashed-dotted line $m/r_0 = m_l/r_0$ in the top panel (a) of Fig.~\ref{fig:mr0}, that is labeled as region I in the bottom panel (b).

\subsubsection{Nonextremal rotating regular black holes with a lightlike boundary at the Cauchy horizon - Line $m=m_l$ }

Configurations with the limiting mass $m_l$, i.e., whose gravitational mass obeys the equality $m/r_0 = m_l/r_0$, present two horizons, a Cauchy horizon and an event horizon. The Cauchy horizon is located exactly at the boundary of the matter distribution, with radius $r_{1-} = r_{1+}=r_0$, while the event horizon is located outside matter. In addition, they present ergosurfaces with the inner ergosurface being located completely inside matter, while the outer ergosurface is located completely outside matter.  Thus, this kind of configurations represents nonextremal rotating regular black holes whose central core of matter presents a lightlike  boundary. 
It is worth mentioning here that this case may be obtained by taking the limit of configurations with a timelike boundary. However, more properly, it can also be obtained by using the smooth junction conditions for a lightlike boundary surface.  The steps of such task are the same as those in Appendix A of Ref.~\cite{Masa22}.  This situation corresponds to the dashed-dotted line $m/r_0 = m_l/r_0$ in panels (a) and (b) of Fig.~\ref{fig:mr0}.

\subsubsection{Nonextremal rotating regular black holes with a timelike boundary - Regions II and IV}

Configurations bearing intermediate masses, i.e., with masses in the interval $m_{c+}/r_0 < m/r_0 < m_l/r_0$, present the two horizons outside matter. The solutions also present two ergosurfaces with the inner ergosurface located inside and outside matter which is located in region II of panel (b) of Fig.~\ref{fig:mr0}, or just outside matter corresponding to region IV of panel (b) of Fig.~\ref{fig:mr0}, and the whole  exterior ergosurface located completely outside matter. The boundary surface of the matter is located inside the Cauchy horizon and, thus, this kind of solutions represents nonextremal rotating regular black holes whose central core of matter has a timelike boundary. This situation corresponds to the region delimited from below by the solid line $m/r_0 = m_{c+}/r_0$ and from above by the dashed-dotted line $m/r_0 = m_{l}/r_0$ of panel (a) of Fig.~\ref{fig:mr0}, that encompasses the two loci labeled as regions II and IV in the bottom panel (b).

\subsubsection{Extremal rotating regular black holes with a timelike  boundary - Line $m=m_{c+}$}

Configurations bearing the extremal mass $m/r_0 = m_{c+}/r_0$ present two coinciding horizons located in the exterior region. The corresponding spacetimes present two ergosurfaces, with the inner ergosurface extending from inside to outside matter for $a/r_0 > 1/3 $ and $m/r_0 > 4/3$, or located just outside matter for $a/r_0 \le 1/3$ and $m/r_0 \le 4/3$, while the outer ergosurface is located outside matter. The boundary of the matter distribution is located inside the degenerate horizon, and the exterior solution is given by the extremal Kerr-Newman geometry, that satisfies the relations $r_{k+} = m = \sqrt{a^2+q^2}$, where $r_{k+}$, with $k=1,\,2$, are the horizon radii. Thus, this kind of solutions represents extremal rotating regular black holes whose central core of matter has a timelike boundary. This situation corresponds to the solid line indicated by the label $m_{c+}/r_0$ in panels (a) and (b) of Fig.~\ref{fig:mr0}.

\subsubsection{Rotating regular star-like objects with a timelike  boundary - Regions III, V, and VI}

Configurations with masses in the interval $0< m/r_0 < m_{c+}/r_0$ present no horizons, although the ergosurfaces can be formed in cases with masses close to the extremal mass $m_{c+}$, and depending on the normalized rotation parameter $a/r_0$. Region III in panel (b) of Fig.~\ref{fig:mr0} corresponds to configurations with ergosurfaces extending from the interior to the exterior spacetime regions. Region V in panel (b) of Fig.~\ref{fig:mr0} corresponds to configurations with ergosurfaces completely outside the matter distribution. Configurations with masses such that $m/r_0 < 1.25$ belong to region VI of panel (b) in Fig.~\ref{fig:mr0}, and correspond to configurations with no ergosurfaces. In contrast to the static solution, these objects with $a\neq 0$ are not necessarily overcharged, since the condition $a^2 + q^2 > m^2$ may be satisfied for sufficiently large $a$, independently of the amount of electric charge. In fact, the electric charge is constrained by the amount of mass through the relation $q^2 = 5m\, r_0/4$, cf. Eq.~\eqref{eq:mr0Rq}. Thus, this kind of configurations represents overextremal rotating regular star-like objects with a timelike boundary, which are represented by the region below the solid line $m/r_0 = m_{c+}/r_0$ in panels (a) and (b) of Fig.~\ref{fig:mr0}, that encompasses the loci labeled as regions III, V, and VI in the bottom panel (b).

\section{Discussion and final comments}
\label{sec:conc}

Static and rotating compact charged regular objects as solutions of the Einstein-Maxwell equations were obtained and studied in the present work. 

In the case of static and spherically symmetric objects, we start by considering that the interior distribution of matter is constituted by a de Sitter-type perfect fluid, in which the pressure $p$ and the energy density $\rho_m$ are related as $p = - \rho_m$, together with the electrified version of the Tolman-like density relation
expressed by $8 \pi \rho_m(r) + q^2(r)\,r^{-4} = 3R^{-2}\left(1 - {r^2}b^{-2}\right)$, where $R$ and $b$ are free parameters. After these assumptions, we are able to solve the Einstein-Maxwell system of equations in order to obtain an exact interior solution that was smoothly matched together with an exterior Reissner-Nordstr\"om spacetime. 
The properties of the complete static solution are analyzed in detail, and it is shown that different kinds of objects emerge in this solution, such as regular nonextremal black holes with a spacelike, a timelike, or a lightlike boundary, 
regular extremal black holes with a timelike boundary, and overcharged tension stars with a timelike boundary.  We also show that by properly taking the limit $b \to \infty$, the L\&Z solution~\cite{Lemos2011} is recovered. Hence, in such a limit, our solution describes also charged regular black holes and overcharged tension stars with an uncharged de Sitter core and a charged boundary shell inside a Reissner-Nordstr\"om region.

All the static regular black hole solutions studied here present the event horizon outside matter. Therefore, all the properties of the Reissner-Nordstr\"om black holes are preserved in our regular models. Such properties include the stability of the event horizon, quasinormal modes, gravitational waves, and black hole thermodynamics.

Moreover, it is known that some regular black hole solutions that possess a Cauchy horizon suffer from the mass inflation instability, similarly to the Reissner-Nordstr\"om black holes. On the other hand, there are other regular black hole solutions that are stable against mass inflation as investigated in Refs.~\cite{Carballo, Bonanno}, even though this topic is still under discussion~\cite{Carballo2023, Bonanno2023}. In this work, we do not investigate the stability against mass inflation of the charged regular black hole solutions. This is an interesting question that we left to a future work.

Additionally, some classes of static compact objects may present stable light rings and, therefore, they may be unstable \cite{Cardoso, Cunha}. A simple analysis shows that all the static charged regular black holes presented in this work, i.e., all the configurations with $0 \le \alpha \le 4/5$ (see panel~(a) of Fig.~\ref{fig:radii}) present only an unstable light ring, that is the same as for the Reissner-Nordstr\"om geometry. On the other hand, the class of static overcharged star-like objects present a stable light ring, together with an unstable light ring, for $4/5 < \alpha < 3\sqrt{2}/5$. For $\alpha = 3\sqrt{2}/5$, this class of overcharged objects presents just one degenerate light ring. Finally, $\alpha > 3\sqrt{2}/5$ the overcharged solutions present no light rings at all. Thus, the location of the inner horizon, outer horizon, stable inner light ring and
unstable outer light ring for different values of $\alpha$ has a similar structure of the Fig.~2 in Ref.~\cite{Rubio}. 

Then, we apply the G\"urses and G\"ursey approach to the non-rotating interior solution mentioned in the last paragraph and construct a charged rotating regular interior, which may be smoothly matched together with the exterior Kerr-Newman spacetime, so that no boundary shell is needed.  The properties of the complete rotating solution are also analyzed in detail and it is shown that different kinds of objects emerge in this solution, such as regular nonextremal black holes with a spacelike, a timelike, or a lightlike boundary, regular extremal black holes with a timelike boundary, and regular undercharged stars with a timelike  boundary. Moreover, it is worth mentioning here that the rotating version of the L\&Z solution will be presented in a separate work.

All the rotating regular black hole solutions obtained in this work present the boundary of the matter core $r_0$ inside the event horizon radius, with the corresponding external geometry being the well-known Kerr-Newman geometry. As a consequence, all properties of such black holes accessible to external observers are preserved in the present model. In particular, observational quantities, such as light rings, gravitational waves, and quasinormal modes, and quantities related to the black hole thermodynamics are exactly the same as in the Kerr-Newman geometry.

It is also known that any rotating configuration that presents an ergoregion but no horizon is unstable to linear perturbations, as argued by Friedman~\cite{Friedman} and numerically investigated for slowly rotating gravastars and boson stars in Ref.~\cite{Cardoso2008}, indicating that ultracompact regular objects with high redshift at their surface are unstable. Hence, some of the present class of charged rotating objects may present ergoregion instabilities. However, in order to have a definite answer to this question, additional investigation is necessary. This and the study of the stability of the central core of matter in the present models of regular compact objects are left for future work.

In summary, we attempted to build a charged rotating regular interior for the Kerr-Newman spacetime, what gave rise to all sorts of compact objects. A possible next step for this work is the study of the maximal analytic extension of all configurations obtained here. In this regard, the extension beyond the disk through the region $r < 0$, for regular compact objects, is still a highly debated topic, see, e.g.~\cite{Ramon, Ramon23, Zhou23}. Such an extension can produce new kinds of objects such as traversable and non-traversable wormholes. 

Another interesting question that this work raises is what kind of electromagnetic properties the interior matter has to possess in order to produce the exterior Kerr-Newman electromagnetic field.  For instance, as investigated by Tiomno~\cite{Tiomno}, a charged rotating oblate ellipsoid, which reproduces the exterior Kerr-Newman electromagnetic field in flat spacetime, allows different types of interior electromagnetic fields with different electromagnetic sources. These issues motivate additional investigation, a task we are pursuing and whose results will be published in a separate work (see Ref.~\cite{Basso:2024hye}).

\begin{acknowledgments}

M.~L.~W.~B.~is funded by Funda\c c\~ao de Amparo \`a Pesquisa do Estado de S\~ao Paulo (FAPESP), Brazil, Grant No.~2022/09496-8. V.~T.~Z.~thanks partial financial support from Conselho Nacional de Desenvolvimento Cien\-t\'ifico
e Tecnol\'ogico (CNPq), Brazil, Grant No.~311726/2022-4, and from Funda\c c\~ao de Aperfei\c coa\-men\-to do Pessoal de N\'ivel Superior (CAPES), Brazil, Grant No. 88887.310351/2018-00.

\end{acknowledgments}

\appendix

\section{The electromagnetic field equations for the rotating spacetime}

\label{sec:appendixA}
Here we analyze the Maxwell equations~\eqref{eq:Max1} and~\eqref{eq:Max2} in the G\"urses-G\"ursey spacetime and show that the electromagnetic field tensor~\eqref{eq:electen} satisfies the Maxwell equations. In order to simplify notation, in what follows we drop the labels $\pm$ that have been used throughout the main text.

The Maxwell equations are given by
\begin{align}
    & \partial_r\left[\left(r^2 +a^2\right)\sin \theta F_{rt}\right] + \partial_{\theta}\left[\sin \theta F_{\theta t}\right] = 4 \pi \sqrt{-g} J^t, \label{eq:mx1} \\
    & \partial_r \left[\csc\theta F_{r \varphi} \right] + \partial_{\theta} \left[\frac{\csc\theta}{r^2 + a^2} F_{\theta \varphi} \right] = - 4 \pi \sqrt{-g} J^{\varphi}, \label{eq:mx2} \\
    & \partial_{r} F_{\theta t} -  \partial_{\theta} F_{rt} = 0, \label{eq:mx3} \\
    & \partial_{r} F_{\theta \varphi} - \partial_{\theta} F_{r \varphi} = 0, \label{eq:mx4}
\end{align}
where $\partial_{\mu} \equiv \partial/\partial x^{\mu}$, $g$ is the determinant of the metric~\eqref{eq:ggmetric}, and $J^{\mu}$ is the current density. 

The non-trivial components of a stationary and axially symmetric electromagnetic tensor are $F_{rt}$, $F_{\theta t}$, $F_{r \varphi}$ and $F_{\theta \varphi}$ which, for the metric~\eqref{eq:ggmetric}, are related
\begin{align}
    F_{r\varphi} = - a \sin^2 \theta F_{rt}, \ a F_{\theta \varphi} = -\left(r^2+a^2\right) F_{\theta t}, \label{eq:a1}
\end{align}
we can rewrite the Maxwell equations in terms only of the components $F_{rt}$ and $F_{\theta t}$, i.e,
\begin{align}
    & \partial_r\left[\left(r^2 +a^2\right)\sin \theta F_{rt}\right] + \partial_{\theta}\left[\sin \theta F_{\theta t}\right] = 4 \pi \sqrt{-g}J^t, \label{eq:mx5} \\
    & \partial_r \left[a \sin \theta F_{r t} \right] + \partial_{\theta} \left[\frac{\csc\theta}{a} F_{\theta t} \right] = 4 \pi\sqrt{-g} J^{\varphi}, \label{eq:mx6} \\
    & \partial_{r} F_{\theta t} -  \partial_{\theta} F_{rt} = 0, \label{eq:mx7} \\
    & \partial_{r}\left[\left(r^2+a^2\right) F_{\theta t}\right] - \partial_{\theta}\left[a^2 \sin^2 \theta F_{r t}\right] = 0. \label{eq:mx8}
\end{align}
Equations \eqref{eq:mx5} and \eqref{eq:mx6} determine the current density inside the charged matter that generates the electromagnetic field given in \eqref{eq:electen}, 
while Eqs.~\eqref{eq:mx7} and~\eqref{eq:mx8} are the compatibility equations. 

From Eq.~\eqref{eq:electen}, it is easy to find the relation
\begin{equation}
    \begin{split}
    \partial_{r} F_{\theta t}  =  \partial_{\theta} F_{rt} = & - \frac{a^2 \sin 2\theta}{\Sigma^3} \Big(r\,Q'(r)\,\Sigma \\
    & - Q(r)\left[3r^2 - a^2 \cos^2 \theta\right]\Big), 
    \end{split}
\end{equation}
which implies that Eq.~\eqref{eq:mx7} is identically satisfied, independently of the charge function $Q(r)$.

The second compatibility condition to be checked is Eq.~\eqref{eq:mx8}. In fact, after using Eq.~\eqref{eq:a1}, it can be rewritten in the form
\begin{align}
    \Sigma \partial_{r} F_{\theta t} = a^2 \sin 2 \theta F_{rt} - 2r F_{\theta t},
\end{align}
and then it is an easy task to verify that the electromagnetic field tensor~\eqref{eq:electen} obeys such a relation, independently of the charge function $Q(r)$. 

The other Maxwell equations, \eqref{eq:mx1} and \eqref{eq:mx2}, furnish the nontrivial components of the current density $J^\mu$, that depends on the charge function $Q(r)$. Such a current density is calculated next.

Given the components of the electromagnetic tensor in Eq.~\eqref{eq:electen}, we have
\begin{align}
     \partial_r F_{rt} = & - \frac{2rQ(r)\left(r^2-3 a^2\cos^2\theta\right)}{\Sigma^3} \nonumber \\
    & + \frac{2Q'(r)(r^2-a^2\cos^2\theta)}{\Sigma^2} - \frac{rQ''(r)}{\Sigma} \nonumber\\
    \partial_{\theta} F_{\theta t} = &\, \frac{2rQ(r)ra^2}{\Sigma^3}\Big(\!\!\left[r^2-3 a^2\cos^2\theta\right]\sin^2 \theta - \Sigma \cos^2 \theta \Big).  \nonumber
\end{align}
These two relations allow us to obtain the components of the current density $J^{\mu}$. After a straightforward but tedious calculation, it follows
\begin{align}
\!\!    4\pi J^t\!& =   \frac{r^2 +a^2}{\Sigma^3} \Big(2Q'(r)\left[r^2-a^2\cos^2\theta\right] - r\,Q''(r)\,\Sigma\Big) \nonumber \\ &\; - \frac{2r^2Q'(r)}{\Sigma^2}\\
 \!\!   4 \pi J^{\varphi} \! &=  \frac{a}{\Sigma^3}\Big(2Q'(r)\left[r^2-a^2\cos^2\theta\right]- r\,Q''(r)\,\Sigma \Big), 
\end{align}
with vanishing $J^r$ and $J^\theta$. The current density $J^{\mu}$ can be decomposed as
\begin{align}
    J^{\mu} = \rho_e u^{\mu} + j^{\mu}, \label{eq:jmu-}
\end{align}
where $u^{\mu} = e_{0}^{\ \mu}$ is the first component of the Carter tetrad (see Eq.~\eqref{eq:rtetrad} in Appendix \ref{sec:appendixEMT}), $\rho_e = - J^{\mu}u_{\mu}$ is charge density as measured by the comoving observer with the fluid and $j^{\mu} = \left(u^{\mu}u_{\nu} + \delta^{\mu}_{\ \nu} \right)J^{\nu}$ is the current density as measure by the comoving observer with fluid. A straightforward calculation shows that
\begin{align}
   \!  4 \pi \rho_{e}& = - \sqrt{\frac{\pm \Delta(r)}{\Sigma}}\left[\frac{2Q'(r)a^2\cos^2\theta}{\Sigma^2} + \frac{r\, Q''(r)}{\Sigma} \right]\!, \label{eq:rhoe-} \\
  \!  4 \pi j^{\mu} & = \frac{2Q'(r)r^2}{\Sigma^2} \frac{a \sin \theta}{\sqrt{\Sigma}} e_3^{\ \mu}, \label{eq:conduction-}
\end{align}
where $ e_3^{\ \mu}$ is the fourth component of the Carter tetrad. Moreover, the projection of the current density $J^{\mu}$ onto the Carter tetrad, i.e., $J^a = e^a_{\ \mu} J^{\mu}$, allows us to express $\rho_e$ and $j^{\mu}$, more succinctly, as $\rho_e = J^0$ and $j^{\mu} = e_3^{\ \mu} J^3 $, where $J^3$ can be extracted from Eq.~\eqref{eq:conduction-}.

\section{The energy-momentum tensor of the rotating geometry}
\label{sec:appendixEMT}

It is well known that the matter source that generates the rotating G\"urses-G\"ursey geometry is an anistropic fluid~\cite{Gurses, Burinskii, Gondolo}, 
whose corresponding energy-momentum tensor can be written in the following form
\begin{align}
    T^{\mu \nu} = \varrho\, e_0^{\ \mu}e_0^{\ \nu} + \mathfrak p_1 e_1^{\ \mu}e_1^{\ \nu} + \mathfrak p_{2} e_2^{\ \mu}e_2^{\ \nu} + \mathfrak p_{3} e_3^{\ \mu}e_3^{\ \nu},\label{eq:ggt}
\end{align}
where 
\begin{align}
    & e_0^{\ \mu} = \frac{1}{\sqrt{\pm \Delta(r) \Sigma}}\Big(\left[r^2 +a^2\right]\delta^{\mu}_{\ t} + a\, \delta^{\mu}_{\ \varphi}\Big), \nonumber \\
    & e_1^{\ \mu} = \sqrt{\frac{\pm \Delta(r)}{\Sigma}} \delta^{\mu}_{\ r}, \ \ e_2^{\ \mu} = \frac{1}{\sqrt{\Sigma}} \delta^{\mu}_{\ \theta}, \label{eq:rtetrad} \\
    & e_3^{\ \mu} =  \frac{1}{\sqrt{\Sigma}\sin \theta}\big[a \sin^2 \theta\delta^{\mu}_{\ t} + \delta^{\mu}_{\ \varphi}\big],\nonumber
\end{align}
is the orthonormal tetrad that diagonalizes the energy-momentum tensor~\eqref{eq:ggt}, also known as the Carter's orthonormal frame. 
The $\pm$ signs in front of $\Delta(r)$ apply to the different regions of the spacetime where $\Delta(r)$ is positive and negative, respectively. 
Thus, the plus sign holds in the regions outside the event horizon and inside the Cauchy horizon, where $\Delta(r)$ is positive and $e_0^{\ \mu}$ is a timelike vector that represents the four-velocity of a stationary observer with angular velocity $\Omega(r) = e_0^{\ \varphi}/e_0^{\ t} = a/(r^2 + a^2)$, while $e_1^{\ \mu}$ is a spacelike vector. The minus sign holds in the region between the event horizon and the Cauchy horizon, where $\Delta(r)$ is negative and the vectors $e_0^{\ \mu}$ and $e_1^{\ \mu}$ exchange roles. Therefore, in this region, $e_1^{\ \mu}$ can be interpreted as the four-velocity of the comoving observer with the fluid. Finally, the elements of the tetrad $e_2^{\ \mu}$ and $e_3^{\ \mu}$ are spacelike  in all regions. 

In order to simplify notation, in what follows we drop the labels $\pm$ that has been used throughout the main text. The eigenvalues $\varrho$, $\mathfrak p_1$, $\mathfrak p_2$ and $\mathfrak p_3$ correspond to the total energy density, the radial, and the tangential pressures of the fluid,
respectively, and are given by~\cite{Burinskii}
\begin{align}
    & \varrho(r,\theta) = -\mathfrak p_1(r,\theta) = \frac{r^2 M'(r)}{4 \pi \Sigma^2}, \label{eq:rpressa}\\
    & \mathfrak p_{2}(r,\theta) = \mathfrak p_{3}(r,\theta) = \frac{r^2 M'(r)}{4 \pi \Sigma^2} - \frac{1}{8 \pi \Sigma}\Big(r\,M(r)\Big)'' \label{eq:tpressa},
\end{align} 
where $M(r)$ is given by Eq.~\eqref{eq:m-(r)}, and the primes indicate derivative with respect to the coordinate $r$. 

We can see that the equations of state are preserved in comparison to the non-rotating original source, namely,  $\varrho = -\mathfrak p_1$ and $\mathfrak p_{1} = \mathfrak p_{3}$. This is in agreement with the fact that, in our case, the Segr\`e type of the  G\"urses and G\"ursey metric in the Kerr-Schild form is $[(11)(1,1)]$, which is the same as the Segr\`e type of the non-rotating metric \eqref{eq:stmetric}, see \cite{Gondolo}. In fact, as shown in \cite{Torres}, the G\"urses-G\"ursey metric \eqref{eq:ggmetric} with a nonconstant mass function $M(r)$ is of Segr\`e type $[(11)(1,1)]$.

Since we are dealing with charged matter, the energy-momentum tensor given by Eq.~\eqref{eq:ggt} may be split into the form $T_{\mu \nu} = M_{\mu \nu} + E_{\mu \nu}$, where $E_{\mu\nu}$ is the energy-momentum tensor of the matter itself, while $M_{\mu\nu}$ is the energy-momentum tensor of the electromagnetic field. In the Carter's orthonormal frame, the nonvanishing components of the electromagnetic energy-momentum tensor are given by
\begin{align}
    & \varrho_{em}(r, \theta) \equiv E_{\mu \nu}e_{0}^{\ \mu} e_{0}^{\ \nu} = \frac{1}{8 \pi}\left(F^2_{rt} + \frac{F^2_{\theta t}}{a^2 \sin^2 \theta} \right), \nonumber\\
    & \mathfrak p_{em 1}(r,\theta) \equiv E_{\mu \nu}e_{1}^{\ \mu} e_{1}^{\ \nu} = - \varrho_{em}(r, \theta),\label{eq:emtEMa}\\
    & \mathfrak p_{em 2}(r,\theta) \equiv E_{\mu \nu}e_{2}^{\ \mu} e_{2}^{\ \nu} = \varrho_{em}(r, \theta), \nonumber \\
    & \mathfrak p_{em 3}(r,\theta) \equiv  E_{\mu \nu}e_{3}^{\ \mu} e_{3}^{\ \nu} = \varrho_{em}(r, \theta). \nonumber
\end{align}

Moreover, the energy-momentum tensor of the matter is determined by the difference between the total energy-momentum tensor and the electromagnetic energy-momentum tensor, 
i.e., $M_{\mu \nu} e_{a}^{\ \mu} e_{b}^{\ \nu} = (T_{\mu \nu} - E_{\mu \nu})e_{a}^{\ \mu} e_{b}^{\ \nu}$. It is straightforward to see that this energy-momentum tensor represents an anisotropic matter fluid with energy density and pressures given by
\begin{align}
    & \varrho_{m}(r, \theta) \equiv M_{\mu \nu} e_{0}^{\ \mu}e_{0}^{\ \nu}= \varrho(r,\theta) - \varrho_{em}(r,\theta) ,\\
    & \mathfrak p_{m 1}(r,\theta) \equiv M_{\mu \nu} e_{1}^{\ \mu} e_{1}^{\ \nu} = -\varrho(r,\theta) +\varrho_{em}(r,\theta),\\
    & \mathfrak p_{m 2}(r,\theta) \equiv M_{\mu \nu} e_{2}^{\ \mu} e_{2}^{\ \nu}=\mathfrak p_2(r,\theta) - \varrho_{em}(r,\theta), \\
    & \mathfrak p_{m 3}(r,\theta) \equiv M_{\mu \nu} e_{3}^{\ \mu} e_{3}^{\ \nu}=\mathfrak p_2(r,\theta) - \varrho_{em}(r,\theta).
\end{align}

\end{document}